\documentclass[aps,prb,twocolumn,superscriptaddress,showpacs,amsmath,amssymb,amsfonts,10pt]{revtex4-1}

\usepackage{amsmath}
\usepackage{amsfonts}
\usepackage{amssymb}
\usepackage{graphicx}
\usepackage{color}
\usepackage{tikz}

\usepackage{lipsum}
\usepackage{dsfont}
\usepackage[caption=false]{subfig}
\usepackage{bm}
\usepackage[normalem]{ulem}

\usepackage[colorlinks,bookmarks=false,urlcolor=blue, citecolor=blue, linkcolor = blue]{hyperref}

\usepackage{pgfplots} 
\usepackage{pgfmath} 
\pgfplotsset{width=0.7\textwidth,compat=1.9} 
\usepgfplotslibrary{fillbetween} 
\pgfkeys{/pgf/number format/.cd,1000 sep={\,}} 
\usetikzlibrary{pgfplots.groupplots} 
\usetikzlibrary{shapes,arrows}

\begin{document}

\title{Shape effects of localized losses in quantum wires: dissipative resonances and  nonequilibrium universality}

\author{Thomas Müller}
\affiliation{Institute for Theoretical Physics, University of Cologne, 50937 Cologne, Germany}
\author{Marcel Gievers}
\affiliation{Max-Planck-Institute  of  Quantum  Optics,  85748 Garching,  Germany}
\affiliation{Munich  Center  for  Quantum  Science  and  Technology,  80799 Munich,  Germany}
\affiliation{Physics Department, Arnold Sommerfeld Center for Theoretical Physics, and Center for NanoScience, Ludwig-Maximilians-Universität München, 80333 Munich, Germany}
\author{Heinrich Fr\"{o}ml}
\affiliation{Institute for Theoretical Physics, University of Cologne, 50937 Cologne, Germany}
\author{Sebastian Diehl}
\affiliation{Institute for Theoretical Physics, University of Cologne, 50937 Cologne, Germany}
\author{Alessio Chiocchetta}
\affiliation{Institute for Theoretical Physics, University of Cologne, 50937 Cologne, Germany}

\begin{abstract}
We study the effects of the spacial structure of localized losses in weakly-interacting fermionic quantum wires. We show that multiple dissipative impurities give rise to resonant effects visible in the transport properties and the particles' momentum distribution. These resonances can enhance or suppress the effective particle losses in the wire. 
Moreover, we investigate the interplay between interactions and the impurity shape and find that, differently from the coherent scatterer case, the impurity shape modifies the scaling of the scattering probabilities close to the Fermi momentum. We show that, while the fluctuation-induced quantum Zeno effect is robust against the shape of the impurities, the fluctuation-induced transparency is lifted continuously. This is reflected in the emergence of a continuous line of fixed points in the renormalization group flow of the scattering probabilities.  
\end{abstract}

\date{\today}

\maketitle

\section{Introduction}
Single-impurity problems are ubiquitous in condensed matter and AMO physics. On the one hand, an impurity can strongly modify the properties of a many-body state,
giving rise, for instance, to the celebrated Kondo effect~\cite{Kondo}, X-ray edge problem~\cite{mahan2013many}, as well as polaron physics~\cite{Devreese_2009,Massignan_2014}. On the other hand, impurities can be used as probes to map out the properties of many-body systems~\cite{Bylander2011,DegenRMP,davis2021}. 
Moreover, the problem of a single degree of freedom interacting with a reservoir is among the simplest problems which exhibit non-trivial many-body effects, and which can be benchmarked with a large number of exact solutions and approximations~\cite{weiss2012quantum,BullaRMP,GullRMP}. 

In one-dimensional systems, impurities play a particularly crucial role, as they may qualitatively alter the transport properties. A remarkable example was given in the seminal paper by Kane and Fisher~\cite{Kane1992PRL,Kane1992PRB,Kane1992PRBRapid}, who showed that the conductance of an electronic quantum wire is dramatically modified by the presence of an impurity: gapless excitations can either enhance or suppress the backscattering due to the impurity. 

Recently, the interest in impurities in one-dimensional quantum systems has been revived by experiments in ultracold atoms, where the impurities were implemented by narrow, highly energetic beams, resulting in an effective localized particle loss~\cite{Barontini2013,Labouvie2016,Muellers2018,Lebrat2019,Corman2019}. As a result, the system is brought out of equilibrium and novel effects arise, prominently the quantum Zeno effect (QZE), whose observation was reported for a Bose gas in a quasi-one-dimensional optical lattice~\cite{Barontini2013,Labouvie2016}, and which was accompanied by a number of theoretical works~\cite{Brazhnyi2009,Shchesnovich2010May,Shchesnovich2010Oct,Witthaut2011,Barmettler2011,Zezyulin2012,Barmettler2011,Kepesidis2012,Kordas2013,Kiefer2017,Kunimi2019,Bychek2019}. Further theoretical investigation unveiled novel effects related to the interplay between localized losses and coherent Hamiltonian dynamics, including the many-body QZE~\cite{Froeml2019,Froeml2020}, dynamical phase transitions~\cite{SELS2020168021,buca2020dissipative,Ueda2021,Haque2020,Saleur2020}, orthogonality catastrophe~\cite{Federico1,Berdanier}, and the engineering of exotic nonequilibrium steady states~\cite{Schnell2017,Lapp2019,Yanay2018,Yanay2020, Krapivsky_2019,Dutta1,Krapivsky_2020,Dutta2,alba2021noninteracting,alba2021unbounded}. Moreover, the quantum Zeno effect has been analyzed also for mobile impurities~\cite{Piazza2021} as well as for dephasing impurities~\cite{Dolgirev,federico2}.

While impurities are often modelled with vanishing width for theoretical convenience (i.e., delta functions), this is not the case in experimental systems, where impurities necessarily have a finite width. Additionally, impurities can also possess a multi-peak structure: for instance, a system of two separated coherent impurities gives rise to resonant tunnelling, which, in the presence of interactions, may lead to an anomalous temperature dependence of the conductance~\cite{Nagaosa1993,Kane1992PRB,Kane1992PRBRapid,Auslaender2000}. However, it has not yet been investigated how the shape of a dissipative impurity impacts the nonequilibrium effects mentioned above.

This article addresses this question by studying different shapes of dissipative impurities in a weakly interacting fermionic quantum wire. The problem of a delta-shaped localized loss was studied in Refs.~\onlinecite{Froeml2019,Froeml2020}: the interplay between gapless quantum modes and a localized impurity was shown to give rise to collective behaviors realizing a dissipative,  nonequilibrium analog of the Kane-Fisher problem~\cite{Kane1992PRL,Kane1992PRB}.
The scattering properties of the impurity were shown to be strongly modified: for repulsive interactions, reflection is enhanced, effectively dividing the wire into two disconnected parts, while for attractive interactions, the localized loss becomes transparent. Within different renormalization group (RG) approaches, this behavior was interpreted in terms of a renormalized dissipation strength $\gamma$ in the vicinity of the Fermi momentum $k_\text{F}$. While the fixed points of the corresponding RG flow are analogous to the Kane-Fisher problem, a qualitatively different approach of the fixed points was found, affecting the physics near the Fermi momentum $k_\text{F}$. In fact, for repulsive interactions, $\gamma$ is infinitely enhanced, and losses are suppressed due to a fluctuation-induced QZE. Instead, for attractive interactions, $\gamma$ vanishes and the backscattering due to the impurity is suppressed, resulting in fluctuation-induced transparency. 
These effects were shown to be visible in the momentum distribution of the particles, exhibiting an anomalous peak close to the Fermi momentum $k_\text{F}$. Moreover, the fluctuation-induced QZE can be detected in transport measurements as a suppression of the transported current, while the fluctuation-induced transparency results in a perfect conductance along the wire~\cite{Marcel}.    
The main results of this article are summarized in the following.

\subsection{Key results}

\emph{(i) Dissipation-induced resonances.} Despite their incoherent nature, we show that multiple dissipative impurities give rise to resonance effects, leading to an oscillatory behavior of the scattering parameters as a function of momentum. While the perfect resonant tunnelling of coherent impurities does not take place, and the transmission probability $\mathcal{T}$ is always smaller than unity, we show that these resonances can either enhance or suppress losses compared to a single dissipative impurity of equal strength.
\\
\emph{(ii) Robustness of the fluctuation-induced QZE.} We demonstrate that, for repulsive interactions, the fixed points of the RG flow equations are insensitive to the shape of the localized loss. This implies that, regardless of the impurity shape, both losses and particle transport are suppressed at the Fermi momentum. Accordingly, the fluctuation-induced QZE is robust against perturbations modifying the shape of a localized loss.\\
\emph{(iii) Novel scaling for the fluctuation-induced transparency.} On the converse, for attractive interactions, we find a novel continuous line of fixed points activated by impurities with a finite width. This implies that the shape of the impurity affects the losses experienced by particles at the Fermi momentum $k_\text{F}$: while vanishing losses are expected for delta-shaped impurities, finite-sized impurities induce a finite amount of losses at $k_\text{F}$. Backscattering, on the converse, remains suppressed. Accordingly, the fluctuation-induced transparency is affected by the shape of a loss in a continuous fashion: small deviations from the perfectly localized case induce weak deviations from perfect transparency.

This article is organized as follows: in Sec.~\ref{Sec:Model} a microscopic model for the fermionic wire with localized dissipation is introduced, while in Sec.~\ref{sec:non-hermitian-scattering} the general scattering properties of dissipative impurities in the absence of interactions are worked out. In Sec.~\ref{sec:ResonantDissipation} resonant effects associated with dissipative impurities, in the absence of interactions, are analyzed. In Sec.~\ref{sec:observables} we discuss experimentally relevant observables. In Sec.~\ref{Sec:RG} we derive RG equations to include the effect of interactions, while in Sec.~\ref{Sec:ScalingGeneric} we analyze the solution of the RG equations. Finally, in Sec.~\ref{sec:conclusion} the main results are summarized and future directions are discussed.

\section{Model and quench protocol}
\label{Sec:Model}

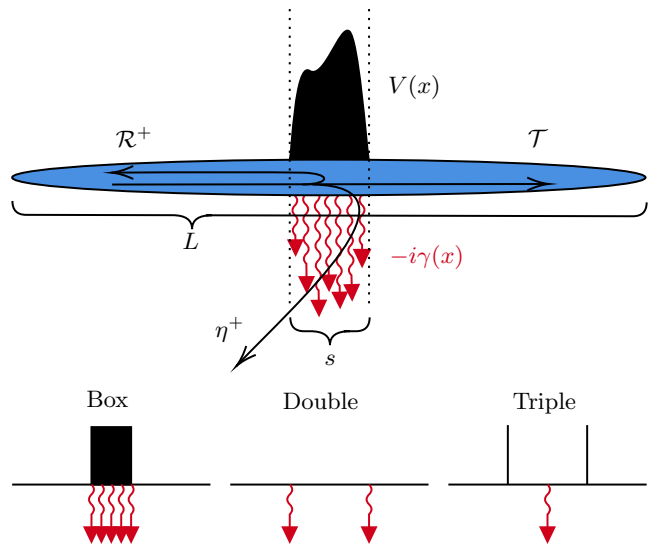
\begin{figure} [t!]

\tikzset{every picture/.style={line width=0.75pt}} 

\begin{tikzpicture}[x=0.75pt,y=0.75pt,yscale=-1,xscale=1]

\draw [fill={rgb, 255:red, 0; green, 0; blue, 0 }  ,fill opacity=1 ]   (140,86.2) .. controls (141.11,81.81) and (142.67,36.37) .. (150,41.2) .. controls (157.33,46.03) and (165.33,17.53) .. (170,21.2) .. controls (174.67,24.87) and (178.52,77.04) .. (179.8,85.73) ;
\draw [color={rgb, 255:red, 208; green, 2; blue, 27 }  ,draw opacity=1 ]   (142.5,104.13) .. controls (144.19,105.77) and (144.22,107.44) .. (142.58,109.13) .. controls (140.94,110.82) and (140.97,112.49) .. (142.66,114.13) .. controls (144.35,115.77) and (144.38,117.44) .. (142.74,119.13) .. controls (141.1,120.82) and (141.13,122.49) .. (142.82,124.13) -- (142.82,124.13) -- (142.95,132.13) ;
\draw [shift={(143,135.13)}, rotate = 269.08] [fill={rgb, 255:red, 208; green, 2; blue, 27 }  ,fill opacity=1 ][line width=0.08]  [draw opacity=0] (8.93,-4.29) -- (0,0) -- (8.93,4.29) -- cycle    ;
\draw [color={rgb, 255:red, 208; green, 2; blue, 27 }  ,draw opacity=1 ][line width=0.75]    (147.9,98.55) .. controls (149.59,100.2) and (149.62,101.86) .. (147.97,103.55) .. controls (146.32,105.24) and (146.34,106.9) .. (148.03,108.55) .. controls (149.72,110.2) and (149.74,111.86) .. (148.09,113.55) .. controls (146.44,115.24) and (146.46,116.9) .. (148.15,118.55) .. controls (149.84,120.2) and (149.86,121.86) .. (148.21,123.55) .. controls (146.56,125.24) and (146.58,126.9) .. (148.27,128.55) .. controls (149.96,130.2) and (149.98,131.86) .. (148.33,133.55) .. controls (146.68,135.24) and (146.7,136.9) .. (148.39,138.54) -- (148.44,142.7) -- (148.54,150.7) ;
\draw [shift={(148.57,153.7)}, rotate = 269.31] [fill={rgb, 255:red, 208; green, 2; blue, 27 }  ,fill opacity=1 ][line width=0.08]  [draw opacity=0] (8.93,-4.29) -- (0,0) -- (8.93,4.29) -- cycle    ;
\draw [color={rgb, 255:red, 208; green, 2; blue, 27 }  ,draw opacity=1 ][line width=0.75]    (153.76,101.98) .. controls (155.45,103.62) and (155.48,105.29) .. (153.84,106.98) .. controls (152.2,108.67) and (152.23,110.34) .. (153.92,111.97) .. controls (155.61,113.61) and (155.64,115.28) .. (154,116.97) .. controls (152.36,118.66) and (152.39,120.33) .. (154.08,121.97) .. controls (155.77,123.61) and (155.8,125.28) .. (154.16,126.97) .. controls (152.52,128.66) and (152.55,130.33) .. (154.24,131.97) .. controls (155.93,133.61) and (155.96,135.28) .. (154.32,136.97) .. controls (152.69,138.66) and (152.72,140.33) .. (154.41,141.97) .. controls (156.1,143.61) and (156.13,145.28) .. (154.49,146.97) .. controls (152.85,148.66) and (152.88,150.33) .. (154.57,151.97) -- (154.61,154.63) -- (154.74,162.63) ;
\draw [shift={(154.79,165.63)}, rotate = 269.08] [fill={rgb, 255:red, 208; green, 2; blue, 27 }  ,fill opacity=1 ][line width=0.08]  [draw opacity=0] (8.93,-4.29) -- (0,0) -- (8.93,4.29) -- cycle    ;
\draw [color={rgb, 255:red, 208; green, 2; blue, 27 }  ,draw opacity=1 ][line width=0.75]    (158.86,100) .. controls (160.55,101.63) and (160.58,103.3) .. (158.95,105) .. controls (157.32,106.7) and (157.35,108.37) .. (159.05,110) .. controls (160.74,111.64) and (160.77,113.31) .. (159.14,115) .. controls (157.51,116.7) and (157.54,118.37) .. (159.24,120) .. controls (160.93,121.64) and (160.96,123.31) .. (159.33,125) .. controls (157.7,126.69) and (157.73,128.36) .. (159.42,129.99) .. controls (161.12,131.62) and (161.15,133.29) .. (159.52,134.99) .. controls (157.89,136.68) and (157.92,138.35) .. (159.61,139.99) -- (159.65,141.92) -- (159.8,149.92) ;
\draw [shift={(159.86,152.92)}, rotate = 268.92] [fill={rgb, 255:red, 208; green, 2; blue, 27 }  ,fill opacity=1 ][line width=0.08]  [draw opacity=0] (8.93,-4.29) -- (0,0) -- (8.93,4.29) -- cycle    ;
\draw [color={rgb, 255:red, 208; green, 2; blue, 27 }  ,draw opacity=1 ][line width=0.75]    (164.05,99.62) .. controls (165.75,101.25) and (165.79,102.91) .. (164.16,104.62) .. controls (162.53,106.32) and (162.57,107.99) .. (164.27,109.62) .. controls (165.97,111.25) and (166.01,112.92) .. (164.38,114.62) .. controls (162.75,116.32) and (162.79,117.98) .. (164.49,119.61) .. controls (166.19,121.24) and (166.23,122.91) .. (164.6,124.61) .. controls (162.97,126.32) and (163.01,127.98) .. (164.72,129.61) .. controls (166.42,131.24) and (166.46,132.91) .. (164.83,134.61) .. controls (163.2,136.31) and (163.24,137.98) .. (164.94,139.61) .. controls (166.64,141.24) and (166.68,142.91) .. (165.05,144.61) .. controls (163.42,146.31) and (163.46,147.98) .. (165.16,149.61) -- (165.18,150.56) -- (165.36,158.56) ;
\draw [shift={(165.43,161.56)}, rotate = 268.72] [fill={rgb, 255:red, 208; green, 2; blue, 27 }  ,fill opacity=1 ][line width=0.08]  [draw opacity=0] (8.93,-4.29) -- (0,0) -- (8.93,4.29) -- cycle    ;
\draw [color={rgb, 255:red, 208; green, 2; blue, 27 }  ,draw opacity=1 ][line width=0.75]    (170,104.29) .. controls (171.71,105.9) and (171.76,107.57) .. (170.14,109.28) .. controls (168.52,110.99) and (168.57,112.66) .. (170.28,114.28) .. controls (171.99,115.9) and (172.04,117.57) .. (170.42,119.28) .. controls (168.8,120.99) and (168.85,122.66) .. (170.56,124.28) .. controls (172.27,125.9) and (172.32,127.57) .. (170.7,129.28) .. controls (169.08,130.99) and (169.13,132.66) .. (170.84,134.27) .. controls (172.55,135.89) and (172.6,137.56) .. (170.98,139.27) .. controls (169.36,140.98) and (169.41,142.65) .. (171.12,144.27) -- (171.19,146.71) -- (171.42,154.7) ;
\draw [shift={(171.5,157.7)}, rotate = 268.39] [fill={rgb, 255:red, 208; green, 2; blue, 27 }  ,fill opacity=1 ][line width=0.08]  [draw opacity=0] (8.93,-4.29) -- (0,0) -- (8.93,4.29) -- cycle    ;
\draw [color={rgb, 255:red, 208; green, 2; blue, 27 }  ,draw opacity=1 ][line width=0.75]    (176.43,98.9) .. controls (178.1,100.57) and (178.1,102.23) .. (176.44,103.9) .. controls (174.78,105.57) and (174.78,107.23) .. (176.45,108.9) .. controls (178.12,110.57) and (178.12,112.23) .. (176.45,113.9) .. controls (174.79,115.57) and (174.79,117.23) .. (176.46,118.9) .. controls (178.13,120.57) and (178.13,122.23) .. (176.47,123.9) .. controls (174.81,125.57) and (174.81,127.23) .. (176.48,128.9) -- (176.48,129.85) -- (176.49,137.85) ;
\draw [shift={(176.5,140.85)}, rotate = 269.9] [fill={rgb, 255:red, 208; green, 2; blue, 27 }  ,fill opacity=1 ][line width=0.08]  [draw opacity=0] (9.82,-4.72) -- (0,0) -- (9.82,4.72) -- cycle    ;
\draw  [fill={rgb, 255:red, 74; green, 144; blue, 226 }  ,fill opacity=1 ] (0,95) .. controls (0,90.03) and (71.51,86) .. (159.71,86) .. controls (247.92,86) and (319.43,90.03) .. (319.43,95) .. controls (319.43,99.97) and (247.92,104) .. (159.71,104) .. controls (71.51,104) and (0,99.97) .. (0,95) -- cycle ;
\draw  [dash pattern={on 0.84pt off 2.51pt}]  (140,10) -- (140,160) ;
\draw  [dash pattern={on 0.84pt off 2.51pt}]  (180,10) -- (180,160) ;
\draw   (140.29,165.61) .. controls (140.29,170.28) and (142.62,172.61) .. (147.29,172.61) -- (150.29,172.61) .. controls (156.96,172.61) and (160.29,174.94) .. (160.29,179.61) .. controls (160.29,174.94) and (163.62,172.61) .. (170.29,172.61)(167.29,172.61) -- (173.29,172.61) .. controls (177.96,172.61) and (180.29,170.28) .. (180.29,165.61) ;
\draw   (0.14,106.86) .. controls (0.13,111.53) and (2.46,113.86) .. (7.13,113.87) -- (80.76,114.04) .. controls (87.43,114.05) and (90.75,116.39) .. (90.74,121.06) .. controls (90.75,116.39) and (94.09,114.07) .. (100.76,114.09)(97.76,114.08) -- (312.98,114.57) .. controls (317.65,114.58) and (319.99,112.25) .. (320,107.58) ;
\draw [color={rgb, 255:red, 0; green, 0; blue, 0 }  ,draw opacity=1 ]   (50.43,98.57) -- (268.71,98.29) ;
\draw [shift={(270.71,98.29)}, rotate = 539.9300000000001] [color={rgb, 255:red, 0; green, 0; blue, 0 }  ,draw opacity=1 ][line width=0.75]    (10.93,-3.29) .. controls (6.95,-1.4) and (3.31,-0.3) .. (0,0) .. controls (3.31,0.3) and (6.95,1.4) .. (10.93,3.29)   ;
\draw [color={rgb, 255:red, 0; green, 0; blue, 0 }  ,draw opacity=1 ]   (146.43,98.57) .. controls (161.29,98.29) and (160.71,92.86) .. (146.43,92.57) .. controls (132.43,92.29) and (83.33,92.32) .. (52.04,92.56) ;
\draw [shift={(50.14,92.57)}, rotate = 359.53999999999996] [color={rgb, 255:red, 0; green, 0; blue, 0 }  ,draw opacity=1 ][line width=0.75]    (10.93,-3.29) .. controls (6.95,-1.4) and (3.31,-0.3) .. (0,0) .. controls (3.31,0.3) and (6.95,1.4) .. (10.93,3.29)   ;
\draw [color={rgb, 255:red, 0; green, 0; blue, 0 }  ,draw opacity=1 ]   (151.57,98.57) .. controls (210.84,99.31) and (142.77,158.87) .. (114.27,188.93) ;
\draw [shift={(113,190.29)}, rotate = 313.08000000000004] [color={rgb, 255:red, 0; green, 0; blue, 0 }  ,draw opacity=1 ][line width=0.75]    (10.93,-3.29) .. controls (6.95,-1.4) and (3.31,-0.3) .. (0,0) .. controls (3.31,0.3) and (6.95,1.4) .. (10.93,3.29)   ;
\draw    (0,250) -- (100,250) ;
\draw    (110,250) -- (210,250) ;
\draw    (220,250) -- (320,250) ;
\draw  [fill={rgb, 255:red, 0; green, 0; blue, 0 }  ,fill opacity=1 ] (40,221) -- (60,221) -- (60,250) -- (40,250) -- cycle ;
\draw [color={rgb, 255:red, 208; green, 2; blue, 27 }  ,draw opacity=1 ]   (40,250) .. controls (41.67,251.67) and (41.67,253.33) .. (40,255) .. controls (38.33,256.67) and (38.33,258.33) .. (40,260) .. controls (41.67,261.67) and (41.67,263.33) .. (40,265) -- (40,269) -- (40,277) ;
\draw [shift={(40,280)}, rotate = 270] [fill={rgb, 255:red, 208; green, 2; blue, 27 }  ,fill opacity=1 ][line width=0.08]  [draw opacity=0] (8.93,-4.29) -- (0,0) -- (8.93,4.29) -- cycle    ;
\draw [color={rgb, 255:red, 208; green, 2; blue, 27 }  ,draw opacity=1 ]   (60,250) .. controls (61.67,251.67) and (61.67,253.33) .. (60,255) .. controls (58.33,256.67) and (58.33,258.33) .. (60,260) .. controls (61.67,261.67) and (61.67,263.33) .. (60,265) -- (60,269) -- (60,277) ;
\draw [shift={(60,280)}, rotate = 270] [fill={rgb, 255:red, 208; green, 2; blue, 27 }  ,fill opacity=1 ][line width=0.08]  [draw opacity=0] (8.93,-4.29) -- (0,0) -- (8.93,4.29) -- cycle    ;
\draw [color={rgb, 255:red, 208; green, 2; blue, 27 }  ,draw opacity=1 ]   (50,250) .. controls (51.67,251.67) and (51.67,253.33) .. (50,255) .. controls (48.33,256.67) and (48.33,258.33) .. (50,260) .. controls (51.67,261.67) and (51.67,263.33) .. (50,265) -- (50,269) -- (50,277) ;
\draw [shift={(50,280)}, rotate = 270] [fill={rgb, 255:red, 208; green, 2; blue, 27 }  ,fill opacity=1 ][line width=0.08]  [draw opacity=0] (8.93,-4.29) -- (0,0) -- (8.93,4.29) -- cycle    ;
\draw [color={rgb, 255:red, 208; green, 2; blue, 27 }  ,draw opacity=1 ]   (55,250) .. controls (56.67,251.67) and (56.67,253.33) .. (55,255) .. controls (53.33,256.67) and (53.33,258.33) .. (55,260) .. controls (56.67,261.67) and (56.67,263.33) .. (55,265) -- (55,269) -- (55,277) ;
\draw [shift={(55,280)}, rotate = 270] [fill={rgb, 255:red, 208; green, 2; blue, 27 }  ,fill opacity=1 ][line width=0.08]  [draw opacity=0] (8.93,-4.29) -- (0,0) -- (8.93,4.29) -- cycle    ;
\draw [color={rgb, 255:red, 208; green, 2; blue, 27 }  ,draw opacity=1 ]   (45,250) .. controls (46.67,251.67) and (46.67,253.33) .. (45,255) .. controls (43.33,256.67) and (43.33,258.33) .. (45,260) .. controls (46.67,261.67) and (46.67,263.33) .. (45,265) -- (45,269) -- (45,277) ;
\draw [shift={(45,280)}, rotate = 270] [fill={rgb, 255:red, 208; green, 2; blue, 27 }  ,fill opacity=1 ][line width=0.08]  [draw opacity=0] (8.93,-4.29) -- (0,0) -- (8.93,4.29) -- cycle    ;
\draw [color={rgb, 255:red, 208; green, 2; blue, 27 }  ,draw opacity=1 ]   (180,250) .. controls (181.67,251.67) and (181.67,253.33) .. (180,255) .. controls (178.33,256.67) and (178.33,258.33) .. (180,260) .. controls (181.67,261.67) and (181.67,263.33) .. (180,265) -- (180,269) -- (180,277) ;
\draw [shift={(180,280)}, rotate = 270] [fill={rgb, 255:red, 208; green, 2; blue, 27 }  ,fill opacity=1 ][line width=0.08]  [draw opacity=0] (8.93,-4.29) -- (0,0) -- (8.93,4.29) -- cycle    ;
\draw [color={rgb, 255:red, 208; green, 2; blue, 27 }  ,draw opacity=1 ]   (140,250) .. controls (141.67,251.67) and (141.67,253.33) .. (140,255) .. controls (138.33,256.67) and (138.33,258.33) .. (140,260) .. controls (141.67,261.67) and (141.67,263.33) .. (140,265) -- (140,269) -- (140,277) ;
\draw [shift={(140,280)}, rotate = 270] [fill={rgb, 255:red, 208; green, 2; blue, 27 }  ,fill opacity=1 ][line width=0.08]  [draw opacity=0] (8.93,-4.29) -- (0,0) -- (8.93,4.29) -- cycle    ;
\draw [color={rgb, 255:red, 208; green, 2; blue, 27 }  ,draw opacity=1 ]   (270,250) .. controls (271.67,251.67) and (271.67,253.33) .. (270,255) .. controls (268.33,256.67) and (268.33,258.33) .. (270,260) .. controls (271.67,261.67) and (271.67,263.33) .. (270,265) -- (270,269) -- (270,277) ;
\draw [shift={(270,280)}, rotate = 270] [fill={rgb, 255:red, 208; green, 2; blue, 27 }  ,fill opacity=1 ][line width=0.08]  [draw opacity=0] (8.93,-4.29) -- (0,0) -- (8.93,4.29) -- cycle    ;
\draw    (250,220) -- (250,250) ;
\draw    (290,220) -- (290,250) ;

\draw (156,183) node [anchor=north west][inner sep=0.75pt]    {$s$};
\draw (84.43,121.36) node [anchor=north west][inner sep=0.75pt]    {$L$};
\draw (188.14,41.71) node [anchor=north west][inner sep=0.75pt]  [color={rgb, 255:red, 0; green, 0; blue, 0 }  ,opacity=1 ]  {$V( x)$};
\draw (189,127.43) node [anchor=north west][inner sep=0.75pt]    {$\textcolor[rgb]{0.82,0.01,0.11}{-i\gamma }\textcolor[rgb]{0.82,0.01,0.11}{(}\textcolor[rgb]{0.82,0.01,0.11}{x}\textcolor[rgb]{0.82,0.01,0.11}{)}$};
\draw (50.86,67.97) node [anchor=north west][inner sep=0.75pt]    {$\mathcal{R}^{+}$};
\draw (258.63,69.39) node [anchor=north west][inner sep=0.75pt]    {$\mathcal{T}$};
\draw (100.94,163.67) node [anchor=north west][inner sep=0.75pt]    {$\eta ^{+}$};
\draw (36,201) node [anchor=north west][inner sep=0.75pt]   [align=left] {Box};
\draw (135,202) node [anchor=north west][inner sep=0.75pt]   [align=left] {Double};
\draw (251,202) node [anchor=north west][inner sep=0.75pt]   [align=left] {Triple};

\end{tikzpicture}

\caption{Upper panel: pictorial representation of a system of one-dimensional spinless fermions with a dissipative impurity of finite width. $\gamma(x)$ is the dissipative impurity profile, while $V(x)$ is the coherent impurity profile. Individual particles coming in from the left may be reflected, transmitted or lost from the system with probabilities $\mathcal{R}^+, \mathcal{T},\eta^+$, respectively.  Lower panel: the three examples of extended dissipative impurities discussed in this article. Black lines represent coherent barriers while wiggled red arrows represent particle losses.}
\label{Fig:Setup}
\end{figure}

We consider a system of spinless fermions with mass $m$ in one dimension, interacting via a short-ranged, translation-invariant interaction potential $g(x)$ as described by the Hamiltonian
\begin{equation} \label{Eq:Hamiltonian}
	\hat{H}_0=\int_x \hat{\psi}^\dagger(x) \frac{-\partial_x^2}{2m} \hat{\psi}(x) + \frac{1}{2} \int_{x,y} g(x-y) \hat{n}(x) \hat{n}(y).
\end{equation}
$\hat{\psi}^\dagger(x),\hat{\psi}(x)$ are fermionic creation and annihilation operators with $\lbrace \hat{\psi}(x),\hat{\psi}^\dagger(y) \rbrace=\delta(x-y)$, $\hat{n}(x)=\hat{\psi}^\dagger(x) \hat{\psi}(x)$ are number operators, $\int_x=\int_{-L/2}^{L/2} dx$ where $L$ is the size of the system, and we set $\hbar=1$ throughout this paper. 
We aim now to model the localized single-particle loss realized in recent experiments~\cite{PhysRevLett.116.235302,PhysRevLett.110.035302,eaat6539,PhysRevLett.123.193605,Corman2019}. In previous theoretical works, a perfectly localized dissipation profile with Markovian single-particle loss was used~\cite{Froeml2019,Froeml2020}. Here, we consider a more realistic description (cf. Fig.~\ref{Fig:Setup}), by allowing for a compact region with particle loss around $x=0$ described by a function $\gamma(x) \geq 0$ extended over the range $-s/2 < x < s/2 $, with $s \ll L$. Outside of this region, the system is translation-invariant, i.e., $\gamma(x)=0$. In addition, we also include a coherent potential $V(x)$ that also extends over $-s/2 < x < s/2 $ with the Hamiltonian $\hat{H}_{\text{imp}}= \int_x V(x) \hat{\psi}^\dagger(x) \hat{\psi}(x)$. The full Hamiltonian then reads $\hat{H}=\hat{H}_0+\hat{H}_{\text{imp}}$.
The time evolution of the system's density matrix is given by the quantum master equation
\begin{equation}
\label{QME}
\partial_t \hat{\rho} = -i [ \hat{H}, \hat{\rho} ] + \mathcal{D}[ \hat{\rho}],
\end{equation}
where $\mathcal{D}$ is the Lindblad superoperator for single-particle losses, following a spatial distribution given by $\gamma(x)$:
\begin{equation}
\label{eq:Lindblad}
	\mathcal{D} [\hat{\rho} ]=  \int_x \gamma(x) \left[ 2 \hat{\psi}(x) \hat{\rho} \hat{\psi}^\dagger(x) - \lbrace \hat{\psi}^\dagger(x) \hat{\psi}(x), \hat{\rho} \rbrace \right].
\end{equation}
We assume that the system is prepared in the ground state of $\hat{H}_0$, with a particle density $n_0$. At $t=0$, the impurities are switched on, and the system approaches a quasi-stationary state, whose lifetime grows linearly with the size of the wire~\cite{Froeml2019,Froeml2020}, and which can be stabilized by adding reservoirs to the ends of the wire~\cite{Marcel}, provided that the reservoirs are Ohmic~\cite{khedri2021luttinger}. This state is characterized by steady particle currents flowing from the far ends of the wire towards the dissipative impurity. In the rest of the paper we focus on this stationary state.

\section{Non-Hermitian scattering problem}
\label{sec:non-hermitian-scattering}
Before including the effect of interactions, we first discuss the general scattering properties of dissipative impurities in the non-interacting case. This fixes the relevant parameter space, describing generic dissipative scattering, which can be exhausted by the various impurity geometries, like box shapes and resonant multi-barrier structures (cf. Fig.~\ref{Fig:Setup}). Delta-shaped impurities are found to be described by a small subset of this parameter space.  \\ 
These scattering properties are naturally encoded in the retarded Green's function, as it describes the response to single-particle perturbations. However, it was shown in Ref.~\onlinecite{Froeml2020} that these properties can be obtained by solving a single-particle Schr\"{o}dinger equation with the non-Hermitian Hamiltonian $\mathcal{H} = -\partial_x^2/(2m)+U(x)$ where $U(x)=V(x)-i \gamma(x)$ (cf. App.~\ref{App:KeldyshFieldTheory}).
The scattering problem associated with the non-Hermitian Hamiltonian can then be solved by using the two ansatzes $\phi_{k}^\pm(x)$:
\begin{subequations}
\label{ScatteringAnsatz}
\begin{align}
\phi_{k}^+(x) & = 
\begin{cases} 		
	e^{i k x} + r_k^+ e^{-ikx} & x < -s/2 \\ 
	t_k^+ e^{ikx} &   x > s/2 
\end{cases},	
 \\
	\phi_k^-(x) & =
\begin{cases}	
	t_k^- e^{-ikx} &  x < -s/2 \\ 
	e^{-ikx}+r_k^- e^{ikx} &   x > s/2
	\end{cases}.
\end{align}
\end{subequations}
$\phi_{k}^\pm(x)$ describe the scattering of a particle coming from the left (+) or right (-) of the impurity, with $r_k^\pm$ and $t_k^\pm$ the corresponding reflection and transmission amplitudes, respectively. These amplitudes fulfill several properties. First, $t_k^+ =t_k^-\equiv t_k$ for any impurity, while $r_k^-= r_k^+$ only for inversion-symmetric impurities, i.e., $U(x) = U(-x)$ (cf. App.~\ref{app:TransferMatrix}). The transmission and reflection probabilities $ \mathcal{T}_k= \vert t_k \vert^2 $ and $\mathcal{R}_k^\pm = \vert r_k^\pm \vert^2$, respectively, satisfy 
\begin{equation}
	\mathcal{T}_k + \mathcal{R}_k^\pm = 1- \eta_k^\pm,
\end{equation}
with $\eta_k^\pm$ the probability of a particle impinging on the impurity from the left (resp. right) to be lost from the system. For a coherent impurity, i.e., $\gamma(x)=0$, the loss probabilities $\eta^\pm_k$ vanish as a consequence of particle-number conservation. 
In the following, we mainly focus on the probabilities $\mathcal{R}^\pm_k,\mathcal{T}_k,\eta_k$, as they directly enter the observables in the stationary state (cf. Sec.~\ref{sec:observables}) and conductance properties~\cite{Marcel}. However, additional information is contained in the phases of $r_k$ and $t_k$ (which do not contribute to  $\mathcal{R}^\pm_k$ and $\mathcal{T}_k$): of particular relevance is their relative phase, which appears in the renormalization of $\mathcal{R}^\pm_k$ and $\mathcal{T}_k$ due to particle interactions (cf. Sec.~\ref{Sec:RG}), and encodes information on the shape of the impurity (cf. Sec.~\ref{Sec:ScalingGeneric}).
By focusing on the inversion-symmetric case, $r_k^+=r_k^-$, in the rest of this section, the information on the relative phase is conveniently encoded in the quantity
\begin{equation} \label{eq:PhaseVar}
	\mathcal{X}_k = \cos \text{arg } t^{*2}_k r_k^2,
\end{equation}
which satisfies $-1 \leq \mathcal{X}_k \leq 1$. In order to determine the possible physical values for the three real parameters 
$\mathcal{R}_k, \mathcal{T}_k, \mathcal{X}_k$, we study the properties of the associated scattering matrix. 
Let $\vec{\phi}_k^\text{in}$ be the two-component vector containing the amplitudes of an ingoing wave, i.e., a wave impinging on the impurity from the left and the right, and $\vec{\phi}_k^\text{out}$ the vector containing the amplitudes of the outgoing wave, i.e., the wave scattered from the impurity. These vectors are related by $\vec{\phi}_k^\text{out}=S_k \vec{\phi}_k^\text{in}$, with $S_k$ the scattering matrix
\begin{equation}
S_k = \left( \begin{array}{cc}
	r_k & t_k \\ 
	t_k & r_k
	\end{array}  \right).
\end{equation}
In a dissipative process, the norm of the state can decrease, and therefore the norm of the outgoing state is smaller or equal to that of the ingoing state, i.e., $||\vec{\phi}_k^\text{out}|| \leq ||\vec{\phi}_k^\text{in}||$. This implies a constraint on the scattering matrix, i.e.,  
\begin{equation} 
\label{CausalityCondition}
 \frac{\vec{\phi}_k^{\text{in} \dagger} S_k^\dagger S_k \vec{\phi}_k^{\text{in}}}{\vec{\phi}_k^{\text{in} \dagger} \vec{\phi}_k^\text{in}} \leq 1.
\end{equation}
For coherent impurities, the equality is satisfied for every $\vec{\phi}_k^\text{in}$ as $S_k^\dagger S_k =\mathds{1}$. For a dissipative impurity, the equality may still be satisfied, although only for certain vectors $\vec{\phi}_k^\text{in}$: these correspond to single-particle dark states of the Lindblad operator~\eqref{eq:Lindblad}. This is the case, for instance, for delta-shaped impurities, i.e.,  $U(x) \propto \delta(x)$, where every wave function with a node at the position of the impurity corresponds to a dark state~\cite{PhysRevA.85.063620}. 
It is worth mentioning the other limiting case of Eq.~\eqref{CausalityCondition}, namely $\vec{\phi}_k^{\text{in} \dagger} S_k^\dagger S_k \vec{\psi}_k^\text{in} = 0$. This corresponds to the so-called coherent perfect absorption, i.e., the extinction of particles impinging on the impurity, and was experimentally observed for a Bose-Einstein condensate with a localized loss~\cite{eaat6539}.  

More generally, we may maximize the left-hand side of Eq.~\eqref{CausalityCondition} to find the smallest possible reduction of a state's norm due to a given dissipative impurity at fixed momentum. This yields the condition (cf. App.~\ref{app:Causality})
\begin{equation}
	2 \mathcal{R}_k \mathcal{T}_k (1+\mathcal{X}_k) \leq \eta_k^2.
	\label{Eq:Constraint}
\end{equation}
\begin{figure}
\centering
\includegraphics[width=\linewidth]{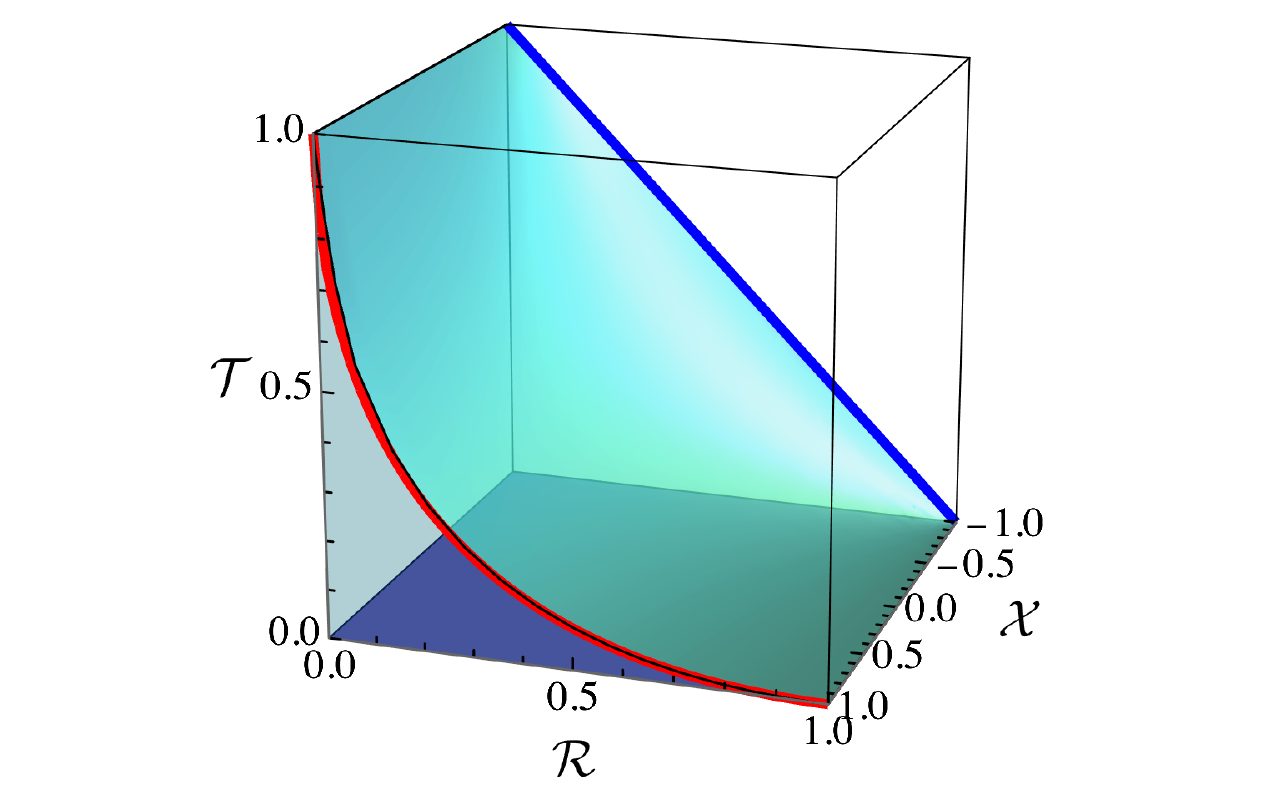}
\caption{The regime of scattering parameters allowed by condition~\eqref{CausalityCondition}: the physical region is enclosed between the plot axes and the surface lying above. Coherent impurities correspond to the blue line, while delta-shaped purely dissipative impurities correspond to the red one. Delta-shaped impurities lie on the surface connecting these lines.}
\label{Fig:ParaRegime}
\end{figure}
Together with $0 \leq \mathcal{R}_k ,\mathcal{T}_k, \eta_k \leq 1$, which holds as these quantities are probabilities, and $-1 \leq \mathcal{X}_k \leq 1$, this defines the physically allowed parameter space as presented in Fig.~\ref{Fig:ParaRegime}. Delta-shaped impurities lie on the surface defined by equality in Eq.~\eqref{Eq:Constraint}. Additional limiting cases are provided by purely dissipative and purely coherent delta-shaped impurities, which correspond to the constraints $\mathcal{X}_k=1$ and $\mathcal{X}_k=-1$, and to the red and blue lines in Fig.~\ref{Fig:ParaRegime}, respectively. 
More generally, for every coherent barrier, $\mathcal{X}_k=-1$ as a consequence of unitarity: accordingly, regardless of the form of the impurity, the transmission and reflection probability always lie on the line $\mathcal{T}_k+\mathcal{R}_k =1$. 
The inequality~\eqref{Eq:Constraint} thus suggests that the shape of a dissipative impurity may play an important role, as it enables to explore the entire three-dimensional manifold allowed by this equation. We substantiate this expectation in Sec.~\ref{Sec:ScalingGeneric}.

A natural question is therefore which minimal model is able to exhaust the entire parameter space defined by the constraint~\eqref{Eq:Constraint}. This can be addressed by a direct inspection of the loss probability, which is constrained to $\eta_k \leq 1/2$ for delta-shaped impurities.

\begin{figure}
\centering
\includegraphics[width=\linewidth]{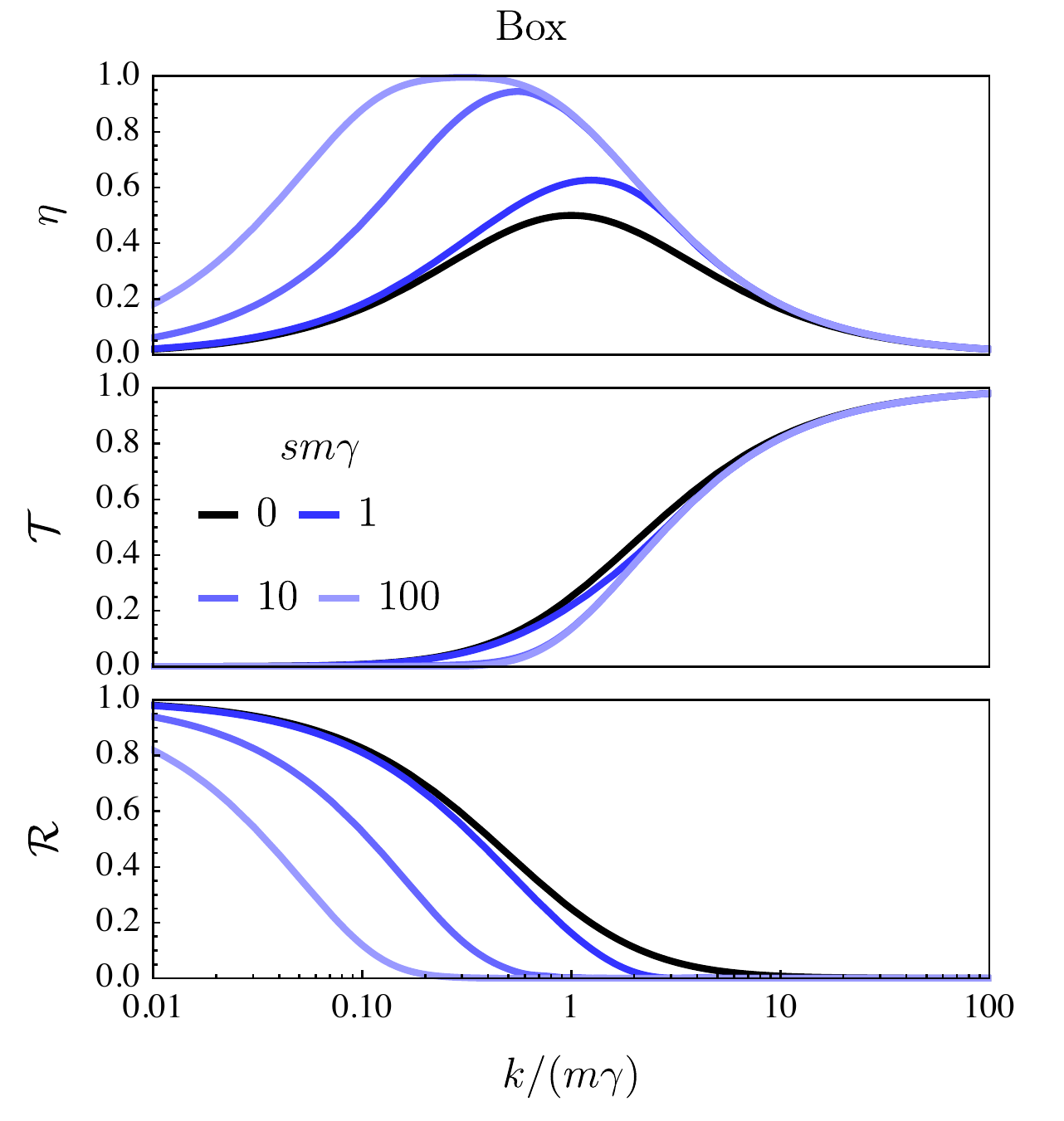}
\caption{Loss ($\eta_k$), transmission ($\mathcal{T}_k$) and reflection ($\mathcal{R}_k$) probabilities for a purely dissipative box-shaped impurity of width $s$ (blue lines), compared to a delta-shaped impurity (black line).}
\label{Fig:BoxLoss}
\end{figure}

We show in the following that a box-shaped impurity can reproduce all the scattering matrices allowed by the inequality~\eqref{CausalityCondition}. To this end, we define the complex potential
\begin{equation}
	U(x) = \begin{cases} U/s & \vert x \vert <s/2 \\ 0 & \text{else} \end{cases},
\end{equation}
with $U=V-i\gamma$, $V \in \mathbb{R}$, and $\gamma>0$. This describes a compact region of length $s$ with spatially constant dissipation and coherent potential. In the limit $s \rightarrow 0$, a delta-shaped impurity with $U(x)=U\delta(x)$ is reproduced. Solving the non-Hermitian Schr\"{o}dinger equation associated with this complex potential yields the loss probability $\eta_k$ displayed in Fig.~\ref{Fig:BoxLoss}. 

For $k \ll m \gamma$ particles are almost completely reflected, while for $k \gg m \gamma$ transmission recovers to one, independently of the value of the impurity width $s$. The loss probability $\eta_k$ vanishes in these two regimes, and has a maximum at intermediate values of $k/ m \gamma$: this non-monotonic behavior is a single-particle incarnation of the quantum Zeno effect, which was already recognized for delta-shaped dissipative impurities~\cite{Froeml2019,Froeml2020}: larger values of the dissipation lead to a counter-intuitive decrease of the particle losses. For $\gamma \to \infty $, the impurity separates the wire into two disconnected, lossless parts.
Remarkably, a finite width $s>0$ enhances the maximum of $\eta_k$, which acquires values larger than the limit $1/2$ for a delta-shaped barrier.

\section{Dissipation and resonance}
\label{sec:ResonantDissipation}

In the presence of two coherent impurities, the transmission displays an oscillatory behavior as a function of the incoming momentum, with the maxima reaching $\mathcal{T}_k=1$, i.e., achieving perfect transmission. These resonances are related to the bound states of the confining potential obtained for infinitely large impurities, and it is known as \emph{resonant tunnelling} in the context of quantum transport~\cite{nazarov_book}, and as \emph{Fabry-Perot resonator} in optics.
Resonant tunnelling also persists in interacting quantum wires, giving rise to a non-monotonic dependence of the conductivity on the temperature or the applied voltage~\cite{Kane1992PRB,Kane1992PRBRapid,Nagaosa1993,Auslaender2000}. A natural question, therefore, concerns the fate of these resonances for dissipative impurities. In this section, we analyze the scattering properties of some physically relevant configurations involving several dissipative impurities, illustrated in Fig.~\ref{Fig:Setup}.

\subsection{Resonant enhancement of losses}

\begin{figure}
\centering
\includegraphics[width=\linewidth]{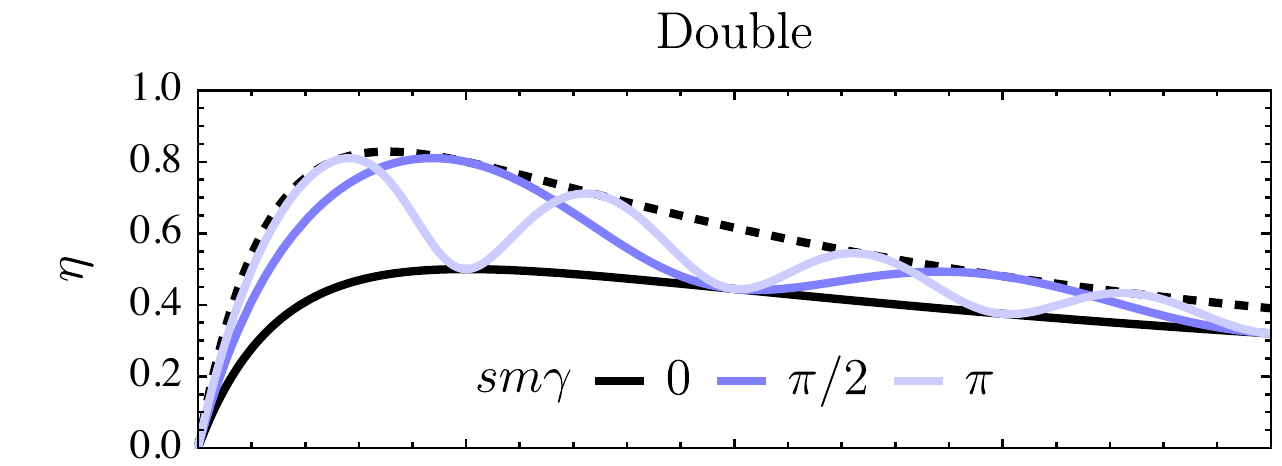}
\includegraphics[width=\linewidth]{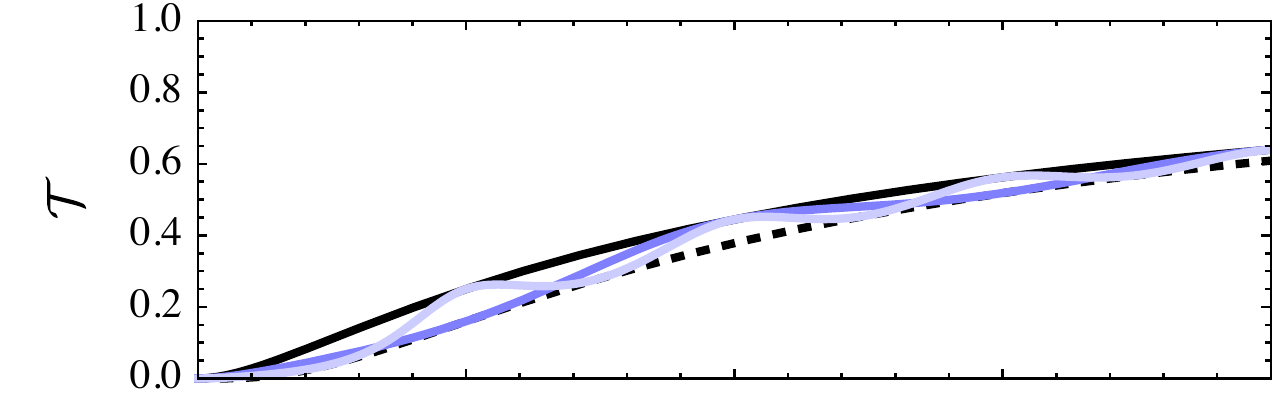}
\includegraphics[width=\linewidth]{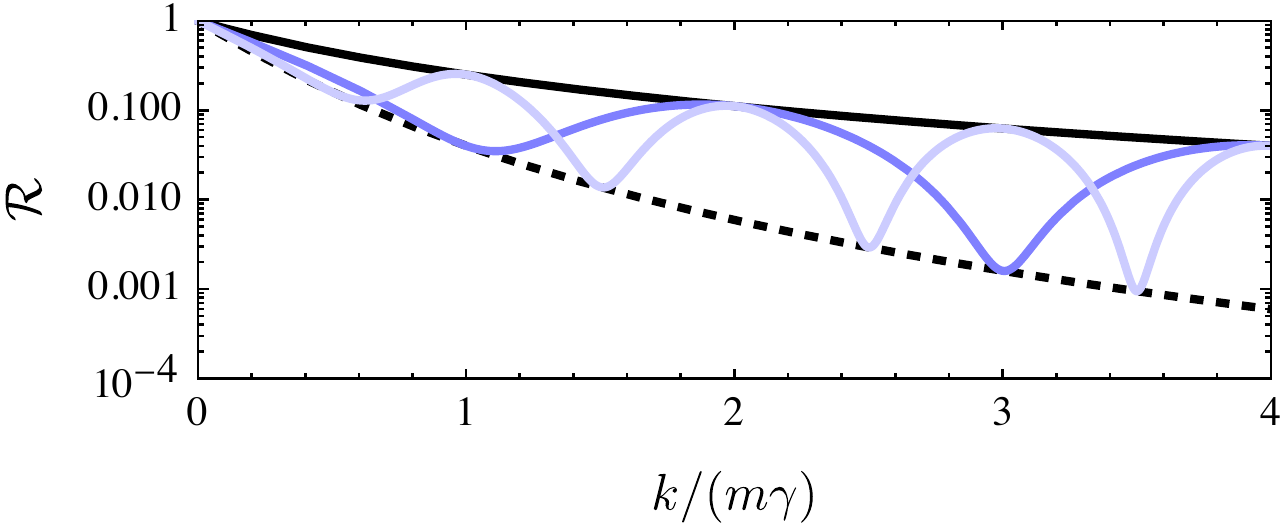}
\includegraphics[width=\linewidth]{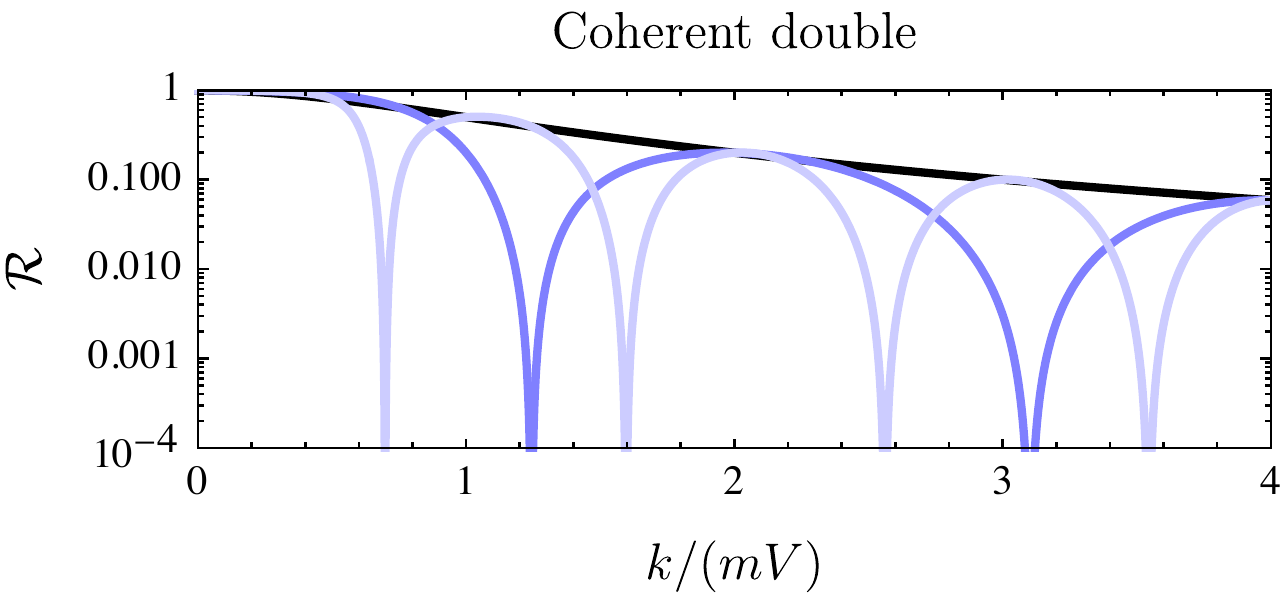}
\caption{
Scattering probabilities for double impurities, as functions of particle momentum $k$ and for different inter-impurity distances $s$. Upper three panels: loss ($\eta_k$), transmission ($\mathcal{T}_k$) and reflection ($\mathcal{R}_k$) probabilities for a double delta-shaped dissipative impurity (blue). The envelope of the functions is given by the scattering parameters of a single delta-shaped impurity (black) and the curve of extremal resonance (dashed), found from maximizing $\eta$ w.r.t $s$.
Lower panel: reflection probability for a double delta-shaped coherent impurity (blue), compared to a single one of equal strength (black).
}
\label{Fig:SymDouble}
\end{figure}

We first consider the non-Hermitian potential given by
\begin{equation}
	U(x) = -\frac{i}{2} \gamma  \bigg[\delta(x-s/2) + \delta(x+s/2)\bigg],
\end{equation}
describing two lossy impurities with identical strength $\gamma/2>0$ separated by a distance $s$. In Fig.~\ref{Fig:SymDouble} we plot the scattering parameters as a function of the momentum for different values of the distance $s$. Compared to a single delta-shaped impurity (black curve), $\eta_k$ may acquire values larger than $1/2$ and, correspondingly, both the reflection and the transmission probabilities are reduced. The most remarkable difference, however, lies in the oscillating behavior as a function of $k$, which testifies the existence of quantum interference caused by the losses. In contrast to the purely coherent case, no perfect transmission is found and, in fact, oscillations are quite small for $\mathcal{T}_k$. On the converse, large oscillations appear for $\mathcal{R}_k$ and $\eta_k$, for which minima and maxima are inverted.
The minima of the oscillations for $\eta_k$ (maxima for $\mathcal{R}_k$) occur for $ks = \pi n$, with $n$ integer: at these points, the scattering parameters acquire the same value as for a single dissipative delta-shaped barrier of strength $\gamma$. The position of the maxima, instead, depends explicitly on the value of $\gamma$. 
The envelope of the maxima can be computed analytically and it is given by $\eta_k=4 \gamma v_k/(v_k^2+ 2\gamma^2 + 2\gamma v_k)$ with $v_k=k/m$, having a maximum of $2 (\sqrt{2}-1)$: accordingly, perfect loss ($\eta_k=1$) cannot be achieved. The presence of additional coherent impurities on top of the dissipative ones does not qualitatively modify this behavior (not shown).

\subsection{Resonant suppression of losses} 
\label{SubSec:resonantsup}
\begin{figure}

\centering
\includegraphics[width=\linewidth]{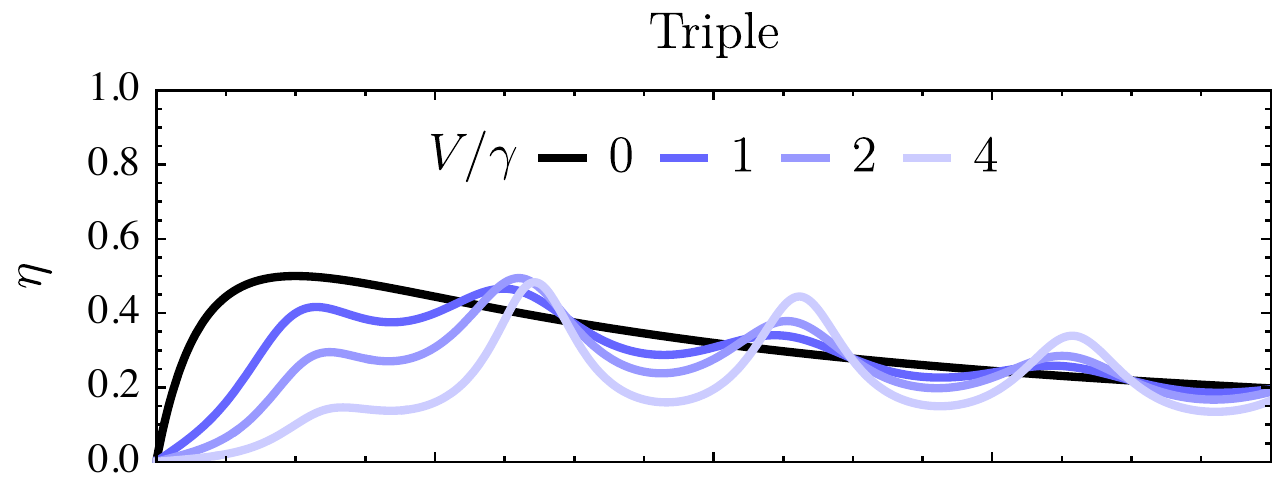}
\includegraphics[width=\linewidth]{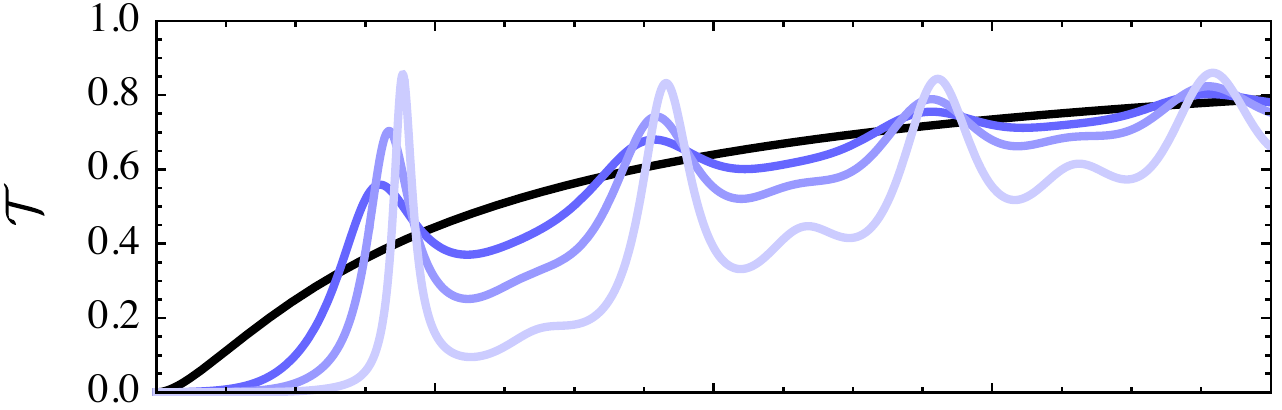}
\includegraphics[width=\linewidth]{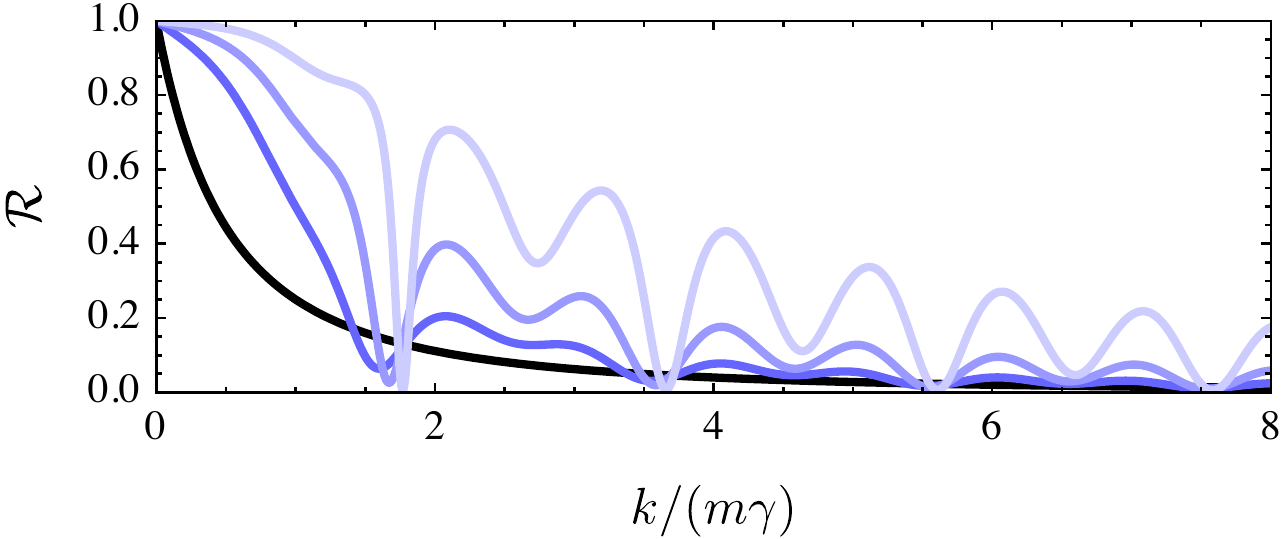}
\caption{
Loss ($\eta_k$), transmission ($\mathcal{T}_k$) and reflection ($\mathcal{R}_k$) probabilities for the impurity~\eqref{Eq:TripleImp} with $s=\pi m \gamma$ as functions of the momentum $k$ and for different values of $V$.}
\label{Fig:triple}
\end{figure}

Although two dissipative impurities cannot give rise to resonant tunneling, the presence of further spatial structure can still enhance transmission at certain momenta. In particular, transmission can be enhanced and losses suppressed by adding two additional coherent barriers. For definiteness, we consider the following structure:
\begin{equation} 
\label{Eq:TripleImp}
	U(x) = \frac{V}{2}\delta(x-s/2)+\frac{V}{2}\delta(x+s/2)-i \gamma \delta(x),
\end{equation}
with $s$ being the distance between the coherent potentials with strength $V$, while $\gamma>0$ is the strength of the dissipation. In Fig.~\ref{Fig:triple} we report the loss, transmission, and reflection probabilities as a function of the momentum $k$ for different values of $V$. A double periodic behavior is observed: $\mathcal{T}_k$ exhibits maxima, whose values increase with $V$, corresponding to the resonances of the double coherent barrier. At these points, reflection is highly suppressed, while losses are  substantially reduced compared to the case with only a dissipative impurity, showing that resonances can be used to reduce the losses caused by a dissipative impurity. 
Besides peaks with enhanced transmission, there exist peaks where the losses are enhanced compared to the single dissipative impurity case.     

Finally, we remark that the effect of an impurity~\eqref{Eq:TripleImp} in a one-dimensional quasi Bose-Einstein condensate was studied in Ref.~\onlinecite{Kunimi2019}, where it was shown to lead to bistable solitonic solutions. 

\section{Observables}
\label{sec:observables}

In this section we show how the scattering properties discussed in the previous section can be directly probed in observables accessible in experiments with ultracold atoms. 
We recall that we are interested in the properties of the system in the quasi-stationary state obtained at large times: accordingly, time-translation invariance is assumed. For a more precise discussion concerning this state, see Refs.~\onlinecite{Froeml2019,Froeml2020}.

A central object is the single-particle equal-time correlation function
\begin{equation} 
\label{Eq:Corrfunc}
	C(x,y) = \langle \hat{\psi}^\dagger(x) \hat{\psi}(y) \rangle,
\end{equation}
which can be evaluated exactly in the absence of interactions by using the initial state and the retarded Green's function (cf. App.~\ref{app:StationaryState} and Ref.~\onlinecite{Froeml2020}). This renders:
\begin{equation}
\label{eq:correlation}
	\!C(x,y) = \!\int_0^{k_\text{F}} \frac{dk}{2\pi}\left[ \phi_k^{+*}(x) \phi_k^{+}(y) + \phi_k^{-*}(x) \phi_k^{-}(y) \right],
\end{equation}
with $\phi_k^{\pm}(x)$ the wave functions defined in Eq.~\eqref{ScatteringAnsatz} and $k_\text{F}$ the Fermi momentum of the initial state. 
Remarkably, the correlations~\eqref{eq:correlation} take the same Fermi-sea structure as in the coherent, equilibrium case~\cite{Matveev1993,Yue1994,Nazarov2003}, upon replacing the single-particle wave functions with the solutions of the corresponding non-Hermitian Schr\"{o}dinger problem. 

\subsection{Particle density}

\begin{figure*}
    \centering
    \includegraphics[width=0.42\linewidth]{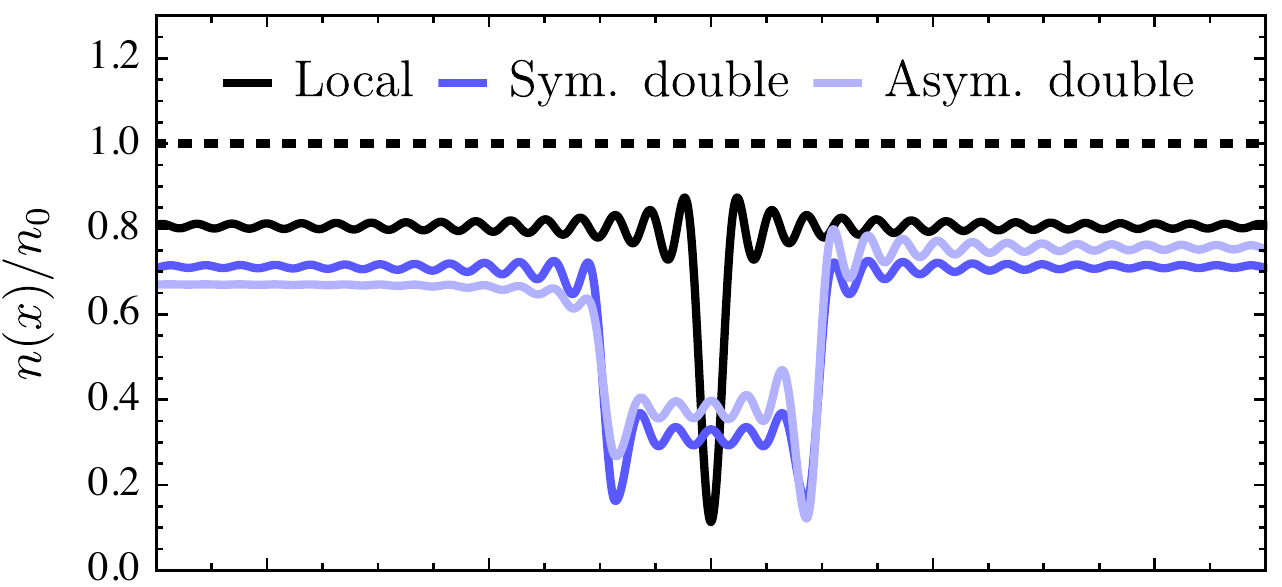} \hspace{10pt}
    \includegraphics[width=0.4275\linewidth]{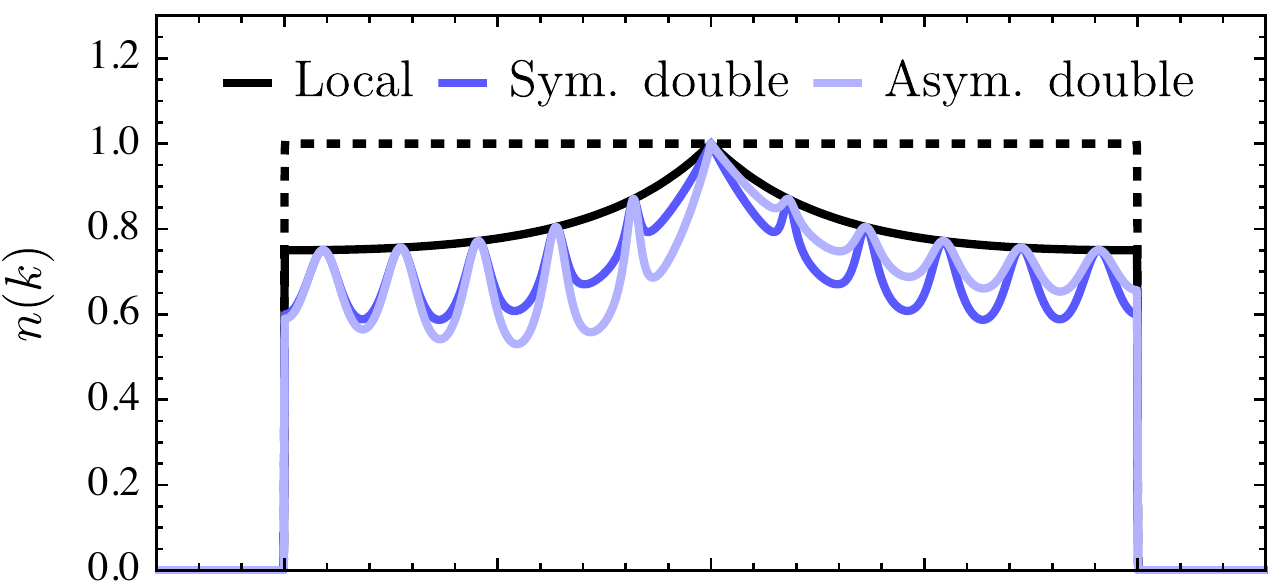}
    \includegraphics[width=0.42\linewidth]{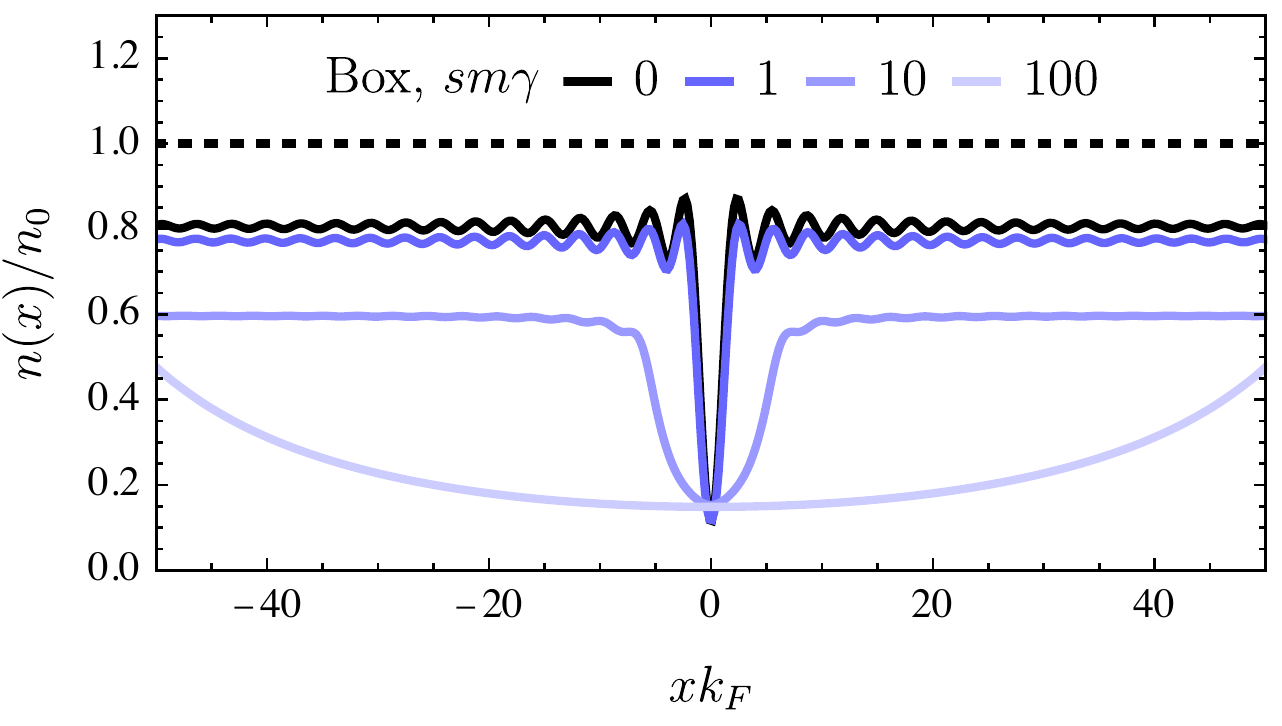} \hspace{10pt}
    \includegraphics[width=0.4275\linewidth]{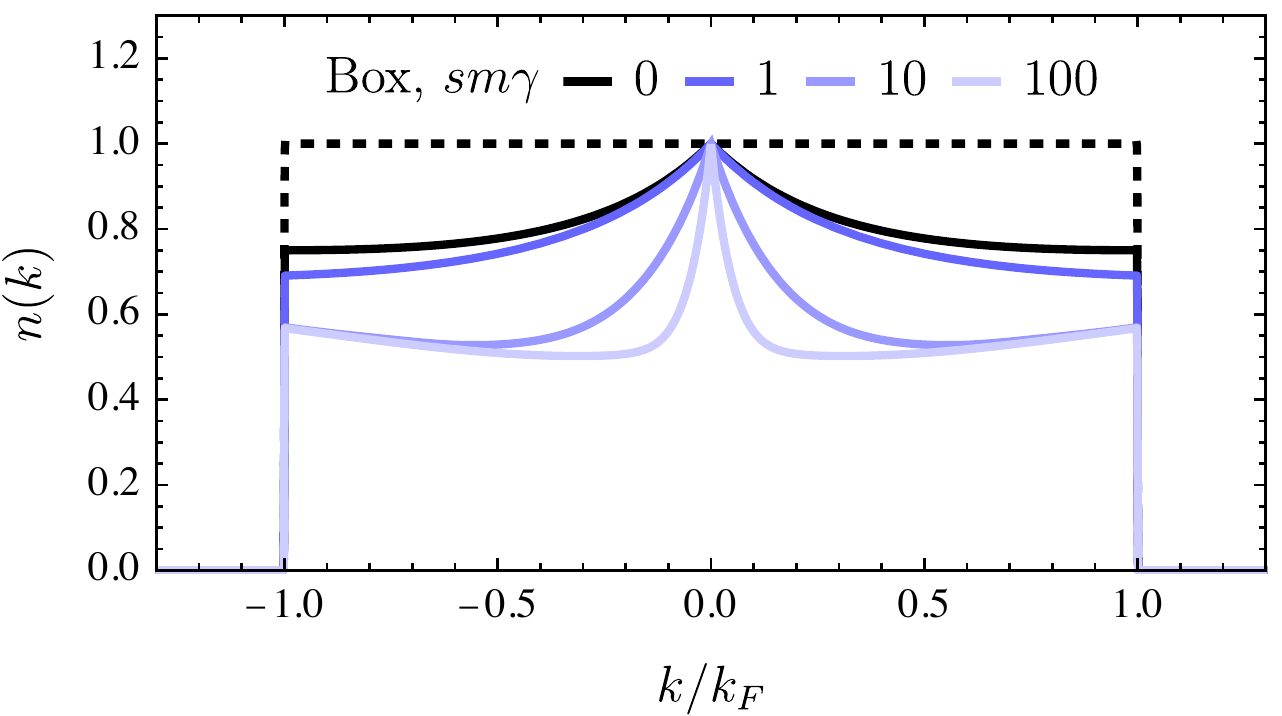}
    \caption{Particle density profile (left panels) and momentum distribution (right panels) for different impurity types. For all the plots $\gamma = k_\text{F}/m$. Dashed lines denote the pre-quench distribution. Upper panels: results for a delta-shaped impurity (black solid line), symmetric double delta-shaped (dark blue line), and asymmetric double delta-shaped impurity (lighter black line). For all curves $s=11\pi/(2m \gamma)$, and $\gamma_\text{L}=\gamma_\text{R}/3$ with $\gamma=\gamma_\text{L}+\gamma_\text{R}$ for the asymmetric double impurity. Lower panels: results for the box-shaped impurity for different values of the impurity width.}
    \label{fig:Densities}
\end{figure*}

The particle density, defined as $n(x) = \langle \hat{\psi}^\dagger(x)\hat{\psi}(x) \rangle$, can be evaluated as $n(x)= C(x,x)$. In the absence of any impurity ($U(x)=0$), $n(x)$ is spatially homogeneous as a consequence of translation invariance, and it takes the initial-state value $n(x)=k_\text{F}/\pi \equiv n_0$, as no quench takes place. For a finite impurity ($U(x)\neq 0$), $n(x)$ is no longer spatially homogeneous: outside of the impurity region it takes the form  
\begin{equation}
	n(x) = \begin{cases}
	\bar{n}^\text{L}+\delta n^\text{L}(x) & x<-s/2 \\ 
	\bar{n}^\text{R} +\delta n^\text{R}(x) & x>s/2
	\end{cases} ,
\end{equation}
with $\delta n^\text{L/R}$ given by
\begin{equation}
		\delta n^\text{L/R}(x) \simeq \frac{\vert r_{k_\text{F}}^\pm \vert \sin \left( 2k_\text{F} \vert x \vert + \text{arg } r_{k_\text{F}}^\pm \right)}{2 \pi \vert x \vert} ,
\end{equation}
for $k_\text{F} |x| \gg 1$ and $x \gg s$. The functions $\delta n^\text{L/R}$ express the Friedel oscillations generated by the impurity, and are a direct consequence of the existence of the persistent Fermi surface in the quasi-stationary state~\cite{Froeml2019,Froeml2020}. These oscillations persist for any impurity shape (cf. Fig~\ref{fig:Densities}).
An important effect of the presence of asymmetric impurities is the existence of background densities $\bar{n}^\text{L/R}$ with different values, namely
\begin{equation}
\label{eq:background_density}
	\bar{n}^\text{L/R} = \frac{1}{\pi} \int_{0}^{k_\text{F}} dk \left( 1- \frac{\eta_k^\pm}{2} \right).
\end{equation}
In fact, non inversion-symmetric dissipative impurities give rise to steady states where the average density is different on the two sides of the impurity (cf. Fig.~\ref{fig:Densities}, upper left panel, lighter solid curve). Notice that a similar imbalanced steady-state was obtained in Ref.~\onlinecite{alba2021noninteracting} for a localized loss supplemented by a localized particle injection. While this effect, combined with interactions may lead to intriguing effects, such as current rectification~\cite{Rect1,Rect2,Rect3}, we do not further investigate its consequences in the present work. 
Accordingly, we restrict ourselves to the case of inversion-symmetric impurities, where $\bar{n}^+=\bar{n}^-$.

\subsection{Momentum distribution}

Another relevant observable is given by the momentum distribution, defined as
\begin{equation} \label{Eq:Mom-Dist-Def}
	n(k) \equiv \frac{1}{L} \int_{x,y} e^{ik(x-y)} \langle \hat{\psi}^\dagger(x)  \hat{\psi}(y) \rangle,
\end{equation}
which can be determined in time-of-flight measurements. By inserting the wave functions~\eqref{ScatteringAnsatz} into Eq.~\eqref{Eq:Corrfunc}, we find (cf. App.~\ref{App:MomentumDistribution})
\begin{equation} 
\label{Eq:Mom-Dist-Res}
	n(k) = \begin{cases} n_\text{F}(k) \left(1 -\eta_{\vert k \vert}^-/2 \right) & k<0 \\ n_\text{F}(k) \left(1 -\eta_{\vert k \vert}^+/2 \right) & k>0 \end{cases} ,
\end{equation}
where $n_\text{F}(k)$ is the momentum distribution in the initial state. %
Eq.~\eqref{Eq:Mom-Dist-Res} has a clear physical interpretation in terms of particles lost from the initial momentum distribution. For instance, a particle impinging on the impurity from the left (i.e., with $k>0$) is lost from the system with the probability $\eta_k^+$. 
The resonances of dissipative impurities directly affect the momentum distribution as shown in Fig.~\ref{fig:Densities}: the oscillating behavior of $\eta_k$ results in a momentum distribution with a peculiar peaked structure.
Moreover, it becomes evident from Eq.~\eqref{Eq:Mom-Dist-Res} that the momentum distribution after the quench preserves the Fermi surface of the initial distribution.
This structure is expected to be robust against the inclusion of interactions, and  visible for times smaller than the thermalization times: this was demonstrated for weak interactions in Ref.~\onlinecite{Froeml2019,Froeml2020}, and numerically for strong interactions in Ref.~\onlinecite{Wolff2020}.
\begin{figure*} 
\centering

\tikzset{every picture/.style={line width=0.75pt}} 

\begin{tikzpicture}[x=0.75pt,y=0.75pt,yscale=-1,xscale=1]

\draw    (226,300) -- (264.59,261.41) ;
\draw [shift={(266,260)}, rotate = 495] [color={rgb, 255:red, 0; green, 0; blue, 0 }  ][line width=0.75]    (10.93,-3.29) .. controls (6.95,-1.4) and (3.31,-0.3) .. (0,0) .. controls (3.31,0.3) and (6.95,1.4) .. (10.93,3.29)   ;
\draw    (266,260) -- (243.44,238.14) ;
\draw [shift={(242,236.75)}, rotate = 404.09000000000003] [color={rgb, 255:red, 0; green, 0; blue, 0 }  ][line width=0.75]    (10.93,-3.29) .. controls (6.95,-1.4) and (3.31,-0.3) .. (0,0) .. controls (3.31,0.3) and (6.95,1.4) .. (10.93,3.29)   ;
\draw   (216,225) .. controls (216,216.72) and (222.72,210) .. (231,210) .. controls (239.28,210) and (246,216.72) .. (246,225) .. controls (246,233.28) and (239.28,240) .. (231,240) .. controls (222.72,240) and (216,233.28) .. (216,225) -- cycle ;
\draw    (242,214.75) -- (264.61,191.44) ;
\draw [shift={(266,190)}, rotate = 494.12] [color={rgb, 255:red, 0; green, 0; blue, 0 }  ][line width=0.75]    (10.93,-3.29) .. controls (6.95,-1.4) and (3.31,-0.3) .. (0,0) .. controls (3.31,0.3) and (6.95,1.4) .. (10.93,3.29)   ;
\draw    (266,190) -- (227.41,151.41) ;
\draw [shift={(226,150)}, rotate = 405] [color={rgb, 255:red, 0; green, 0; blue, 0 }  ][line width=0.75]    (10.93,-3.29) .. controls (6.95,-1.4) and (3.31,-0.3) .. (0,0) .. controls (3.31,0.3) and (6.95,1.4) .. (10.93,3.29)   ;
\draw  [dash pattern={on 4.5pt off 4.5pt}]  (266,150) -- (266,300) ;
\draw    (312,300) -- (350.59,261.41) ;
\draw [shift={(352,260)}, rotate = 495] [color={rgb, 255:red, 0; green, 0; blue, 0 }  ][line width=0.75]    (10.93,-3.29) .. controls (6.95,-1.4) and (3.31,-0.3) .. (0,0) .. controls (3.31,0.3) and (6.95,1.4) .. (10.93,3.29)   ;
\draw  [dash pattern={on 4.5pt off 4.5pt}]  (352,150) -- (352,300) ;
\draw    (352,260) -- (375.59,236.58) ;
\draw [shift={(377.01,235.17)}, rotate = 495.2] [color={rgb, 255:red, 0; green, 0; blue, 0 }  ][line width=0.75]    (10.93,-3.29) .. controls (6.95,-1.4) and (3.31,-0.3) .. (0,0) .. controls (3.31,0.3) and (6.95,1.4) .. (10.93,3.29)   ;
\draw    (377.01,214.17) -- (353.44,191.39) ;
\draw [shift={(352,190)}, rotate = 404.02] [color={rgb, 255:red, 0; green, 0; blue, 0 }  ][line width=0.75]    (10.93,-3.29) .. controls (6.95,-1.4) and (3.31,-0.3) .. (0,0) .. controls (3.31,0.3) and (6.95,1.4) .. (10.93,3.29)   ;
\draw    (352,190) -- (313.41,151.41) ;
\draw [shift={(312,150)}, rotate = 405] [color={rgb, 255:red, 0; green, 0; blue, 0 }  ][line width=0.75]    (10.93,-3.29) .. controls (6.95,-1.4) and (3.31,-0.3) .. (0,0) .. controls (3.31,0.3) and (6.95,1.4) .. (10.93,3.29)   ;
\draw  [dash pattern={on 4.5pt off 4.5pt}]  (65.5,150) -- (65.5,300)(62.5,150) -- (62.5,300) ;
\draw    (24.41,266.17) -- (62.61,226.44) ;
\draw [shift={(64,225)}, rotate = 493.88] [color={rgb, 255:red, 0; green, 0; blue, 0 }  ][line width=0.75]    (10.93,-3.29) .. controls (6.95,-1.4) and (3.31,-0.3) .. (0,0) .. controls (3.31,0.3) and (6.95,1.4) .. (10.93,3.29)   ;
\draw    (64,225) -- (25.82,186.58) ;
\draw [shift={(24.41,185.17)}, rotate = 405.18] [color={rgb, 255:red, 0; green, 0; blue, 0 }  ][line width=0.75]    (10.93,-3.29) .. controls (6.95,-1.4) and (3.31,-0.3) .. (0,0) .. controls (3.31,0.3) and (6.95,1.4) .. (10.93,3.29)   ;
\draw  [dash pattern={on 4.5pt off 4.5pt}]  (165.18,150) -- (165.18,300) ;
\draw    (125,265.67) -- (163.77,226.42) ;
\draw [shift={(165.18,225)}, rotate = 494.65] [color={rgb, 255:red, 0; green, 0; blue, 0 }  ][line width=0.75]    (10.93,-3.29) .. controls (6.95,-1.4) and (3.31,-0.3) .. (0,0) .. controls (3.31,0.3) and (6.95,1.4) .. (10.93,3.29)   ;
\draw    (165.18,225) -- (126.91,186.58) ;
\draw [shift={(125.5,185.17)}, rotate = 405.11] [color={rgb, 255:red, 0; green, 0; blue, 0 }  ][line width=0.75]    (10.93,-3.29) .. controls (6.95,-1.4) and (3.31,-0.3) .. (0,0) .. controls (3.31,0.3) and (6.95,1.4) .. (10.93,3.29)   ;
\draw   (372,225) .. controls (372,216.72) and (378.72,210) .. (387,210) .. controls (395.28,210) and (402,216.72) .. (402,225) .. controls (402,233.28) and (395.28,240) .. (387,240) .. controls (378.72,240) and (372,233.28) .. (372,225) -- cycle ;
\draw  [dash pattern={on 4.5pt off 4.5pt}]  (500,150) -- (500,300) ;
\draw    (440,260) -- (463.41,236.58) ;
\draw [shift={(464.82,235.17)}, rotate = 494.99] [color={rgb, 255:red, 0; green, 0; blue, 0 }  ][line width=0.75]    (10.93,-3.29) .. controls (6.95,-1.4) and (3.31,-0.3) .. (0,0) .. controls (3.31,0.3) and (6.95,1.4) .. (10.93,3.29)   ;
\draw    (463.82,214.17) -- (441.4,191.42) ;
\draw [shift={(440,190)}, rotate = 405.40999999999997] [color={rgb, 255:red, 0; green, 0; blue, 0 }  ][line width=0.75]    (10.93,-3.29) .. controls (6.95,-1.4) and (3.31,-0.3) .. (0,0) .. controls (3.31,0.3) and (6.95,1.4) .. (10.93,3.29)   ;
\draw   (460.19,224.91) .. controls (460.19,216.67) and (466.86,210) .. (475.09,210) .. controls (483.33,210) and (490,216.67) .. (490,224.91) .. controls (490,233.14) and (483.33,239.81) .. (475.09,239.81) .. controls (466.86,239.81) and (460.19,233.14) .. (460.19,224.91) -- cycle ;

\draw (219.5,215.07) node [anchor=north west][inner sep=0.75pt]    {$I^{+*}_{k}$};
\draw (83.5,216.4) node [anchor=north west][inner sep=0.75pt]    {$=$};
\draw (186.5,216.4) node [anchor=north west][inner sep=0.75pt]    {$+$};
\draw (292.5,216.4) node [anchor=north west][inner sep=0.75pt]    {$+$};
\draw (376,215.07) node [anchor=north west][inner sep=0.75pt]    {$I^{-*}_{k}$};
\draw (418,216.4) node [anchor=north west][inner sep=0.75pt]    {$+$};
\draw (460.32,215.07) node [anchor=north west][inner sep=0.75pt]    {$-I^{+}_{k}$};
\draw (20,214.4) node [anchor=north west][inner sep=0.75pt]   {$r^{+}_{k,\text{eff}}$};
\draw (136.68,213.9) node [anchor=north west][inner sep=0.75pt]    {$r^{+}_{k}$};
\draw (238.5,249.9) node [anchor=north west][inner sep=0.75pt]    {$r^{+}_{k}$};
\draw (238.5,179.4) node [anchor=north west][inner sep=0.75pt]    {$r^{+}_{k}$};
\draw (336,243.9) node [anchor=north west][inner sep=0.75pt]    {$t_{k}$};
\draw (355.5,174.4) node [anchor=north west][inner sep=0.75pt]    {$t_{k}$};
\draw (161.18,132.4) node [anchor=north west][inner sep=0.75pt]    {$U$};
\draw (262,132.4) node [anchor=north west][inner sep=0.75pt]    {$U$};
\draw (348,132.4) node [anchor=north west][inner sep=0.75pt]    {$U$};
\draw (496,132.4) node [anchor=north west][inner sep=0.75pt]    {$U$};
\draw (53,131.9) node [anchor=north west][inner sep=0.75pt]    {$U_{\text{eff}}$};

\end{tikzpicture}

\vspace{0.5cm}

\tikzset{every picture/.style={line width=0.75pt}} 

\begin{tikzpicture}[x=0.75pt,y=0.75pt,yscale=-1,xscale=1]

\draw  [dash pattern={on 4.5pt off 4.5pt}]  (62.31,150) -- (62.31,300)(59.31,150) -- (59.31,300) ;
\draw    (21.23,266.17) -- (59.43,226.44) ;
\draw [shift={(60.81,225)}, rotate = 493.88] [color={rgb, 255:red, 0; green, 0; blue, 0 }  ][line width=0.75]    (10.93,-3.29) .. controls (6.95,-1.4) and (3.31,-0.3) .. (0,0) .. controls (3.31,0.3) and (6.95,1.4) .. (10.93,3.29)   ;
\draw    (60.81,225) -- (98.6,186.59) ;
\draw [shift={(100,185.17)}, rotate = 494.53] [color={rgb, 255:red, 0; green, 0; blue, 0 }  ][line width=0.75]    (10.93,-3.29) .. controls (6.95,-1.4) and (3.31,-0.3) .. (0,0) .. controls (3.31,0.3) and (6.95,1.4) .. (10.93,3.29)   ;
\draw  [dash pattern={on 4.5pt off 4.5pt}]  (295,150) -- (295,300) ;
\draw    (255,300) -- (293.59,261.41) ;
\draw [shift={(295,260)}, rotate = 495] [color={rgb, 255:red, 0; green, 0; blue, 0 }  ][line width=0.75]    (10.93,-3.29) .. controls (6.95,-1.4) and (3.31,-0.3) .. (0,0) .. controls (3.31,0.3) and (6.95,1.4) .. (10.93,3.29)   ;
\draw    (271.19,214.17) -- (293.6,191.42) ;
\draw [shift={(295,190)}, rotate = 494.58] [color={rgb, 255:red, 0; green, 0; blue, 0 }  ][line width=0.75]    (10.93,-3.29) .. controls (6.95,-1.4) and (3.31,-0.3) .. (0,0) .. controls (3.31,0.3) and (6.95,1.4) .. (10.93,3.29)   ;
\draw   (245,225) .. controls (245,216.72) and (251.72,210) .. (260,210) .. controls (268.28,210) and (275,216.72) .. (275,225) .. controls (275,233.28) and (268.28,240) .. (260,240) .. controls (251.72,240) and (245,233.28) .. (245,225) -- cycle ;
\draw    (295,260) -- (272.09,236.6) ;
\draw [shift={(270.69,235.17)}, rotate = 405.61] [color={rgb, 255:red, 0; green, 0; blue, 0 }  ][line width=0.75]    (10.93,-3.29) .. controls (6.95,-1.4) and (3.31,-0.3) .. (0,0) .. controls (3.31,0.3) and (6.95,1.4) .. (10.93,3.29)   ;
\draw    (295,190) -- (333.59,151.41) ;
\draw [shift={(335,150)}, rotate = 495] [color={rgb, 255:red, 0; green, 0; blue, 0 }  ][line width=0.75]    (10.93,-3.29) .. controls (6.95,-1.4) and (3.31,-0.3) .. (0,0) .. controls (3.31,0.3) and (6.95,1.4) .. (10.93,3.29)   ;
\draw  [dash pattern={on 4.5pt off 4.5pt}]  (397,150) -- (397,300) ;
\draw    (357,300) -- (395.59,261.41) ;
\draw [shift={(397,260)}, rotate = 495] [color={rgb, 255:red, 0; green, 0; blue, 0 }  ][line width=0.75]    (10.93,-3.29) .. controls (6.95,-1.4) and (3.31,-0.3) .. (0,0) .. controls (3.31,0.3) and (6.95,1.4) .. (10.93,3.29)   ;
\draw    (397,260) -- (420.26,237.11) ;
\draw [shift={(421.69,235.71)}, rotate = 495.47] [color={rgb, 255:red, 0; green, 0; blue, 0 }  ][line width=0.75]    (10.93,-3.29) .. controls (6.95,-1.4) and (3.31,-0.3) .. (0,0) .. controls (3.31,0.3) and (6.95,1.4) .. (10.93,3.29)   ;
\draw    (421.69,214.21) -- (398.43,191.4) ;
\draw [shift={(397,190)}, rotate = 404.44] [color={rgb, 255:red, 0; green, 0; blue, 0 }  ][line width=0.75]    (10.93,-3.29) .. controls (6.95,-1.4) and (3.31,-0.3) .. (0,0) .. controls (3.31,0.3) and (6.95,1.4) .. (10.93,3.29)   ;
\draw    (397,190) -- (435.59,151.41) ;
\draw [shift={(437,150)}, rotate = 495] [color={rgb, 255:red, 0; green, 0; blue, 0 }  ][line width=0.75]    (10.93,-3.29) .. controls (6.95,-1.4) and (3.31,-0.3) .. (0,0) .. controls (3.31,0.3) and (6.95,1.4) .. (10.93,3.29)   ;
\draw   (417,225) .. controls (417,216.72) and (423.72,210) .. (432,210) .. controls (440.28,210) and (447,216.72) .. (447,225) .. controls (447,233.28) and (440.28,240) .. (432,240) .. controls (423.72,240) and (417,233.28) .. (417,225) -- cycle ;
\draw  [dash pattern={on 4.5pt off 4.5pt}]  (171,150) -- (171,300) ;
\draw    (131.33,266) -- (169.61,226.44) ;
\draw [shift={(171,225)}, rotate = 494.05] [color={rgb, 255:red, 0; green, 0; blue, 0 }  ][line width=0.75]    (10.93,-3.29) .. controls (6.95,-1.4) and (3.31,-0.3) .. (0,0) .. controls (3.31,0.3) and (6.95,1.4) .. (10.93,3.29)   ;
\draw    (171,225) -- (209.25,186.75) ;
\draw [shift={(210.67,185.33)}, rotate = 495] [color={rgb, 255:red, 0; green, 0; blue, 0 }  ][line width=0.75]    (10.93,-3.29) .. controls (6.95,-1.4) and (3.31,-0.3) .. (0,0) .. controls (3.31,0.3) and (6.95,1.4) .. (10.93,3.29)   ;

\draw (28.5,213.5) node [anchor=north west][inner sep=0.75pt]    {$t_{k,\text{eff}}$};
\draw (102,212) node [anchor=north west][inner sep=0.75pt]    {$=$};
\draw (269.5,248) node [anchor=north west][inner sep=0.75pt]    {$r^{+}_{k}$};
\draw (249.5,214.31) node [anchor=north west][inner sep=0.75pt]    {$I^{+*}_{k}$};
\draw (333,215.81) node [anchor=north west][inner sep=0.75pt]    {$+$};
\draw (389.5,132) node [anchor=north west][inner sep=0.75pt]    {$U$};
\draw (48.5,131.5) node [anchor=north west][inner sep=0.75pt]    {$U_\text{eff}$};
\draw (287,131) node [anchor=north west][inner sep=0.75pt]    {$U$};
\draw (421.5,215) node [anchor=north west][inner sep=0.75pt]    {$I^{-*}_{k}$};
\draw (406,177.5) node [anchor=north west][inner sep=0.75pt]    {$r^{-}_{k}$};
\draw (280,170.5) node [anchor=north west][inner sep=0.75pt]    {$t_{k}$};
\draw (380.5,241) node [anchor=north west][inner sep=0.75pt]    {$t_{k}$};
\draw (163,132) node [anchor=north west][inner sep=0.75pt]    {$U$};
\draw (153.67,208.33) node [anchor=north west][inner sep=0.75pt]    {$t_{k}$};
\draw (221.33,215.33) node [anchor=north west][inner sep=0.75pt]    {$+$};

\end{tikzpicture}
\caption{Symbolic representation of the first order perturbation theory for the scattering parameters $r_k^+$ and $t_k$. Due to interactions, an effective impurity potential $U_{\text{eff}}(x)$ is seen by the particles instead of the bare impurity potential $U(x)$. To first order in the interactions, only the presented processes, involving a single scattering from the Friedel oscillations, contribute to the correction of the scattering parameters. Such a scattering corresponds to a divergent integral $I_k^\pm$ in Eq.~\eqref{Eq:divergentCorrection}.}
\label{Fig:ReflectionSeries}
\end{figure*}
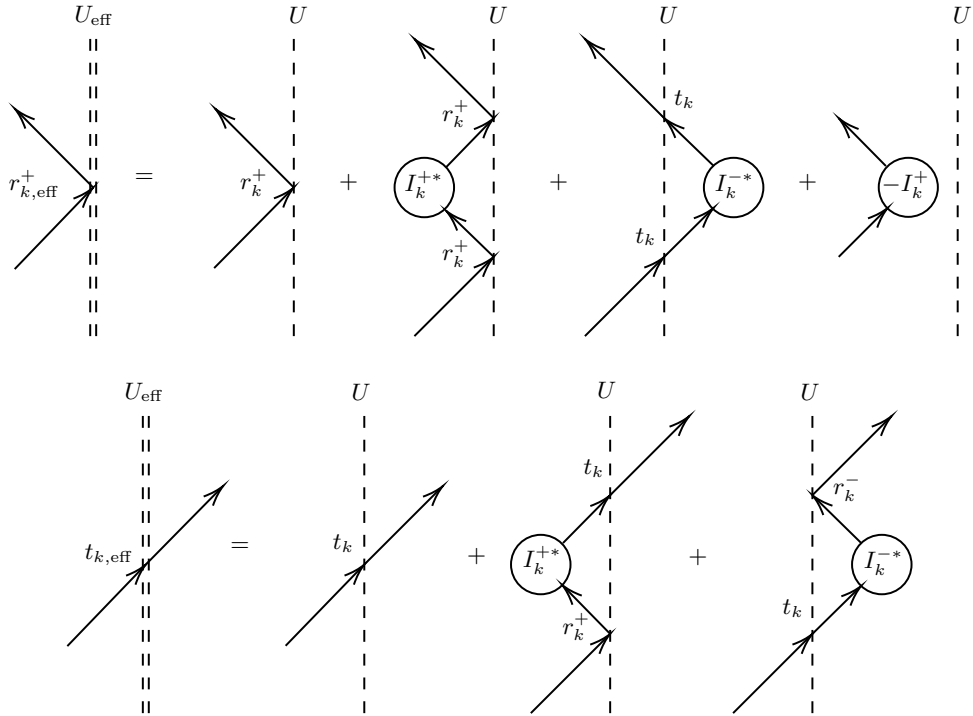

\section{Effect of weak interactions: renormalization group analysis}
\label{Sec:RG}
In this section we analyze the effect of weak interactions on the scattering properties of impurities and, consequently, on the observables discussed in Sec.\ref{sec:observables}. It was shown that weak interactions strongly renormalize the transport close to the Fermi momentum for both coherent~\cite{Kane1992PRL,Kane1992PRB,Matveev1993,Yue1994} and dissipative impurities~\cite{Froeml2019,Froeml2020}. This effect was interpreted as a consequence of the effective single-particle scattering potential generated by the Friedel oscillations dressed by interactions. 
To this end, we derive the RG flow equations for the scattering parameters for impurities of arbitrary shape, following Refs.~\onlinecite{Matveev1993,Yue1994,Nazarov2003,Froeml2019,Froeml2020}.

\subsection{Perturbation theory}

As a first step, we compute the corrections $\delta \phi_k^\pm(x)$ to the wave functions~\eqref{ScatteringAnsatz} to first order in the interaction g, which yields~\cite{Froeml2019,Froeml2020}
\begin{multline}
\delta \phi_k^\pm(x) = \int_{y,z}G_\text{R}(k^2/(2m),x,y) g(y-z) \\
\times \bigg[ n(z) \phi_k^\pm(y) - C(y,x)\phi_k^\pm(z)\bigg],
\end{multline}
where $G_\text{R}^{-1}= -i\partial_t +i0^+ + \partial_x^2/(2m) -U(x)$ is the retarded Green's function of the non-interacting problem (cf. App.~\ref{App:KeldyshFieldTheory}), $g(x)$ is the inter-particle interaction (cf. Eq.~\eqref{Eq:Hamiltonian}), while $n(x)$ and $C(x,y)$ are the non-interacting particle density and correlations discussed in Sec.~\ref{sec:observables}. 
Far away from the impurity, the corrections $\delta \phi_k^\pm(x)$ due to the Friedel oscillations acquire the same scattering form as in Eq.~\eqref{ScatteringAnsatz}, with modified reflection and transmission amplitudes. The corrections to $r_k$ and $t_k$ read: 
\begin{subequations} 
\label{Eq:divergentCorrection}
\begin{align}
	\delta r_k^\pm &= \frac{\alpha}{2} \left( r_k^\pm I_k^{\pm*} r_k^\pm + t_k I_k^{\mp *} t_k - I_k^\pm \right), \\
	\delta t_k &= \frac{\alpha}{2} \left( t_k I_k^{+*} r_k^+ + r_k^- I_k^{-*} t_k \right),
\end{align}
\end{subequations}
where $\alpha = [\tilde{g}(0)-\tilde{g}(2k_\text{F})]/2\pi v_\text{F}$, with $v_\text{F} \equiv k_\text{F}/m $ the Fermi velocity and $\tilde{g}(k)$ the Fourier transform of the interaction $g(x)$, quantifies the strength of the interaction, with $\alpha>0$ ($\alpha<0$) for repulsive (attractive) ones, and 
\begin{equation}
\label{eq:divergingIntegral}
	I_k^\pm = \int_0^{k_\text{F}} dp \frac{r_p^\pm}{p+i0^+-k}.
\end{equation}
This integral is logarithmically divergent for $k \rightarrow k_\text{F}$, as a consequence of the resonance of the wave function with the Friedel oscillations.
The form of the corrections in Eq.~\eqref{Eq:divergentCorrection} can be interpreted as a  sum over all the possible scattering processes involving a single scattering off a Friedel oscillation (cf. Fig.~\ref{Fig:ReflectionSeries}).

Finally, we note that for an inversion-symmetric impurity, the constant part $\bar{n}^\text{L}=\bar{n}^\text{R}$ acts only as a constant potential shift and can be removed by shifting all energies. Breaking this symmetry invalidates this argument, so that we assume $r_k^+=r_k^-=r_k$ for the rest of the paper.

\subsection{Renormalization group}

The divergence of the integral in Eq.~\eqref{eq:divergingIntegral} spoils the validity of the perturbative analysis, which then requires a more refined treatment. Given its logarithmic character, a natural approach involves its resummation using a renormalization group treatment: this was pioneered in Refs.~\onlinecite{Matveev1993, Yue1994,Nazarov2003} for coherent impurities, and subsequently extended in Refs.~\onlinecite{Froeml2019,Froeml2020} for dissipative ones. This results in RG equations whose solution provides a resummation of the logarithmic divergences. Moreover, from the solution of the RG equations it is possible to reconstruct the dependence of the scattering coefficients on the momentum $k$, system size $L$, or the initial temperature $T$.
The RG equations for $r_k$ and $t_k$ are given by:
\begin{subequations}
\begin{align}
	\partial_\ell r_k &= \frac{\alpha}{2} \left[ r_k^* \left( r_k^2 + t_k^2 \right) - r_k \right], \\
	\partial_\ell t_k &= \alpha t_k \vert r_k \vert^2,
\end{align}
\end{subequations}
where $\ell$ is the dimensionless RG flow parameter, which is given by $\ell = -\log(\Delta/\Delta_\text{UV})$, with $\Delta=\text{min }(\vert k-k_\text{F} \vert,1/L,T/v_\text{F})$ and $\Delta_\text{UV}$ a microscopic energy scale.
The scattering parameters of the non-interacting problem provide the initial values for the RG flow at $\ell=0$. This allows us to treat the momentum dependence due to the impurity shape and due to the interactions simultaneously. 

\section{Solution of the RG equations}
\label{Sec:ScalingGeneric}

\subsection{General structure}

\begin{figure} 
\includegraphics[width=\linewidth]{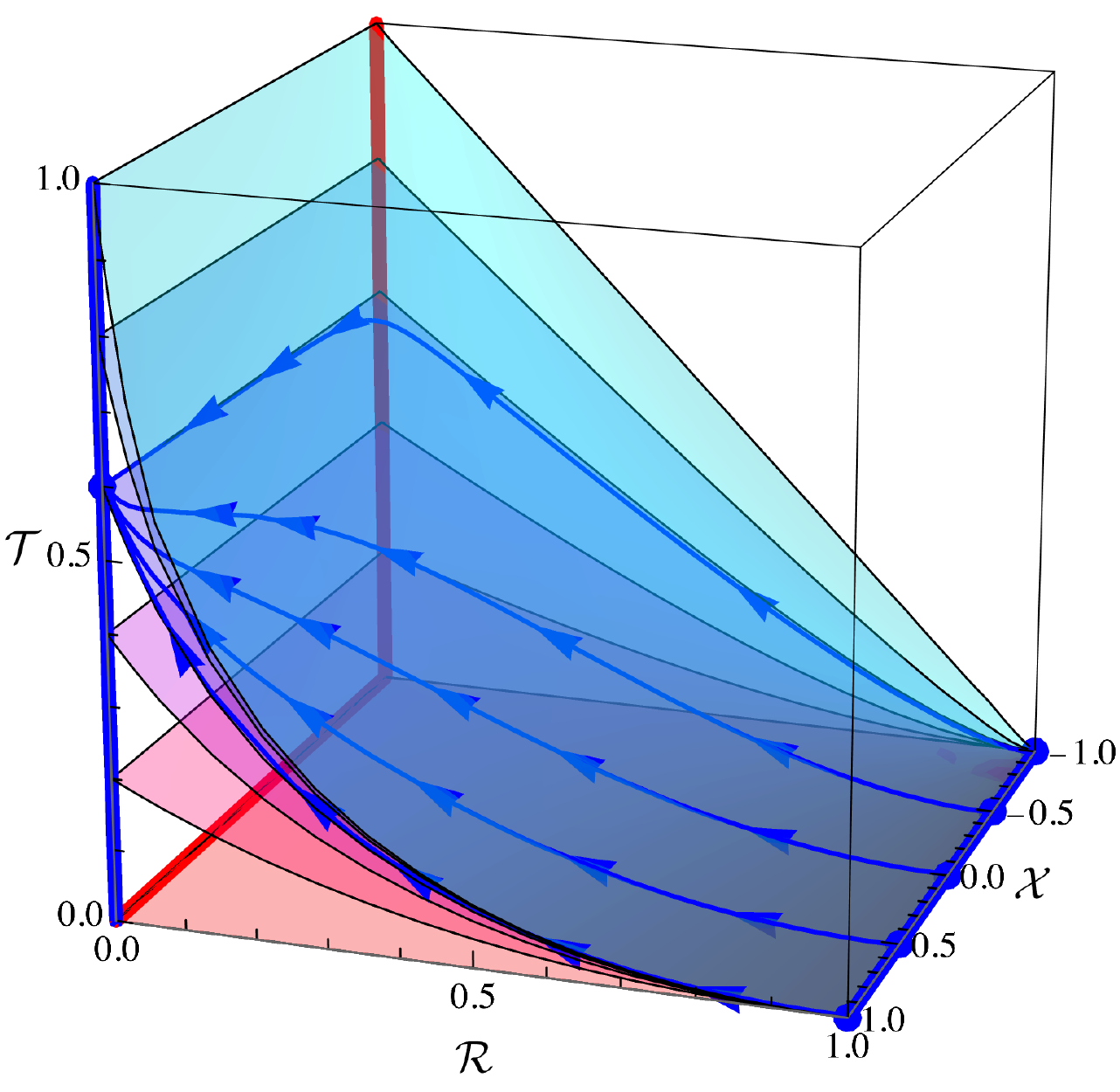}
\caption{
RG flow for attractive interactions ($\alpha <0$) and $\eta^\star=0.4$ (arrows). Surfaces with constant $\mathcal{D}>0$ are shown for $\eta^\star=(0,0.2,0.4,0.6,0.8,1)$. Blue, thick lines correspond to fixed points stable either for $\alpha>0$ or $\alpha<0$. The red thick lines are unstable fixed points of the RG for both repulsive and attractive interactions.}
\label{3DFLowPlot}
\end{figure}

\begin{figure}
	\includegraphics[width=\linewidth]{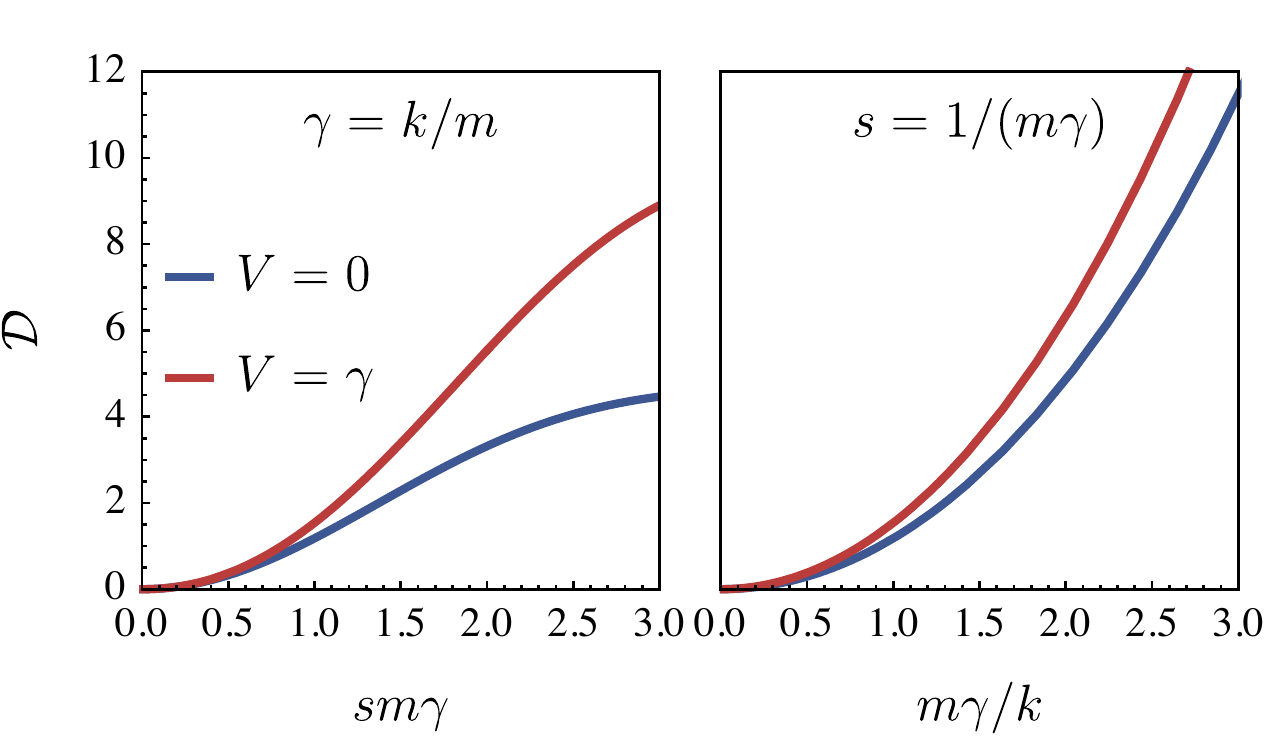}
	\caption{Conserved quantity $\mathcal{D}$ for the box-shaped impurity. Both a finite width and a finite loss are required for $\mathcal{D}>0$ while coherent potentials weakly modify it additionally. For $s \rightarrow 0$, we find $\mathcal{D} \sim s^2 \gamma^2$.}
	\label{Fig:Ddistance}
\end{figure}

We begin with a discussion of the general RG flow for bare scattering parameters discussed in Sec.~\ref{sec:non-hermitian-scattering}, realized for instance by box-shaped impurities (cf. Fig.~\ref{Fig:BoxLoss}). Note that the phase of $t_k$ is not renormalized, $\partial_\ell \text{arg } t_k=0$, such that it is convenient to recast the RG equations in terms of three real parameters, namely the transmission probability $\mathcal{T}_k = \vert t_k \vert^2$, the reflection probability $\mathcal{R}_k = \vert r_k \vert^2$ and the parameter encoding the relative phase between transmission and reflection amplitudes, i.e.,  $\mathcal{X}_k = \cos \text{arg }(t_k^{*2} r_k^2)$ (cf. Eq.~\eqref{eq:PhaseVar}). 
We omit the momentum index in the following and find 
\begin{subequations} \label{Eq:ScalGen}
\begin{align}
	\partial_\ell \mathcal{R} &= -\alpha \mathcal{R} \left( \mathcal{R} -1+\mathcal{TX} \right), \\
	\partial_\ell \mathcal{T} &= -2 \alpha \mathcal{RT}, \\
	\partial_\ell \mathcal{X} &= -\alpha \mathcal{T} \left( 1-\mathcal{X}^2 \right).
\end{align}
\end{subequations}
For a visualization of these equations see Fig.~\ref{3DFLowPlot}. The planes with $\mathcal{R}=0$, $\mathcal{T}=0$, $\mathcal{X}=\pm 1$, as well as $\eta^2=2\mathcal{RT}(1+\mathcal{X})$ are separatrices of the RG flow, so that the physical constraints on bare impurities, as discussed in Sec.~\ref{sec:non-hermitian-scattering}, cannot be violated due to the RG flow. This implies that the picture of scattering from an effective impurity is still justified in the presence of interactions as there always exists a bare complex impurity with the same set of parameters. A crucial feature of the RG equations~\eqref{Eq:ScalGen} is the existence of a quantity $\mathcal{D}$ conserved by the RG flow, namely 
\begin{equation} 
\label{eq:Ddefinition}
\mathcal{D} \equiv  \frac{\eta^2-2\mathcal{RT}(1+\mathcal{X})}{\mathcal{T}} \geq 0,
\end{equation}
which is related to the width of the dissipative impurity (cf. Fig.~\ref{Fig:Ddistance}).
The immediate consequence of this conserved quantity is that the RG flow is constrained onto a two-dimensional manifold in the $(\mathcal{R},\mathcal{T},\mathcal{X})$ space for a given initial condition. While for purely coherent or delta-shaped impurities $\mathcal{D}=0$, $\mathcal{D}$ is finite for generic impurities. The value of $\mathcal{D}$ for a box-shaped impurity is shown in Fig.~\ref{Fig:Ddistance} for a fixed value of $k$: it increases upon increasing the width $s$ or the dissipation strength $\gamma$. Finally, the equations admit four different lines of fixed points (FPL) given by:
\begin{subequations}
\begin{align}
&\text{FPL}_1: \mathcal{R}^\star=0, \mathcal{X}^\star = 1, \\
&\text{FPL}_2: \mathcal{R}^\star=0, \mathcal{X}^\star = -1, \\
&\text{FPL}_3: \mathcal{R}^\star=1, \mathcal{T}^\star = 0, \\
&\text{FPL}_4: \mathcal{R}^\star=0, \mathcal{T}^\star = 0. 
\end{align}
\end{subequations}
In the following, we analyze the stability of these fixed points and their implication for the observables.
We concentrate on stable fixed points as they describe the scattering properties for particles with momentum $k \rightarrow k_\text{F}$ in the limit $L \rightarrow \infty, T \rightarrow 0$.
\subsection{Stable fixed points for repulsive interactions}

For repulsive interactions ($\alpha>0$) the stable line of fixed points is given by FPL$_3$, which is characterized by arbitrary values of $\mathcal{X}$ and perfect reflection $\mathcal{R}=1$. This moreover implies that $\eta=0$ at the fixed point: modes close to the Fermi momentum are thus dissipationless, corresponding to the fluctuation-induced quantum Zeno effect~\cite{Froeml2019,Froeml2020}. 
In order to quantify the stability properties, we linearize the RG equations~\eqref{Eq:ScalGen} by writing $\mathcal{T} = \mathcal{T}^\star+\delta \mathcal{T}$, $\mathcal{R} = \mathcal{R}^\star+\delta \mathcal{R}$ and $\mathcal{X} = \mathcal{X}^\star+\delta \mathcal{X}$, and retaining only the linear order in the distance from the fixed point. The resulting equations are given by:
\begin{equation}
	\partial_\ell
	\begin{pmatrix}
	\delta\mathcal{R} \\ 
	\delta\mathcal{T} \\ 
	\delta \mathcal{X}
	\end{pmatrix}
	=
	 -\alpha 
	 \begin{pmatrix}
	1 & \mathcal{X}^\star & 0 \\ 
	0 & 2 & 0 \\ 
	0 & 1-\mathcal{X}^\star & 0
	\end{pmatrix}
	\begin{pmatrix}
	\delta\mathcal{R} \\ 
	\delta\mathcal{T} \\ 
	\delta\mathcal{X}
	\end{pmatrix}.
\end{equation}
The eigenvalues of the matrix denote stable, unstable and marginal directions depending on the sign of the eigenvalues (negative, positive, zero, respectively). In the present case, we find the eigenvalues $-\alpha ,-2 \alpha$ and $0$, indicating two stable directions and a marginal one, which corresponds to the line of fixed points itself. 
Further insight on the properties of the fixed point can be gained by inspecting the scaling of $\delta\mathcal{T}$ and $\delta\mathcal{R}$:
\begin{subequations}
	\begin{align}
		\delta\mathcal{R} &= \mathcal{X}^\star \delta\mathcal{T}_0 e^{-2 \alpha \ell} - \left( \mathcal{X}^\star \delta\mathcal{T}_0 - \delta\mathcal{R}_0\right) e^{-\alpha \ell} ,\\
		\delta\mathcal{T} &= \delta\mathcal{T}_0 e^{-2\alpha \ell}, 
	\end{align}
\end{subequations}
with $\delta\mathcal{T}_0, \delta\mathcal{R}_0$ the initial displacement from the fixed point values.
While the scaling of $\delta\mathcal{T}$ with $\ell$ depends only on $\alpha$, the one for $\delta\mathcal{R}$ depends additionally on the specific values of 
$\mathcal{X}^\star, \delta\mathcal{T}_0, \delta\mathcal{R}_0$. In particular, while generically the leading scaling is given by $\sim e^{-\alpha \ell}$, it turns to $e^{-2 \alpha \ell}$ if $\mathcal{X}^\star \delta\mathcal{T}_0 = \delta\mathcal{R}_0$. 
This condition corresponds precisely to purely coherent impurities, where $\mathcal{X}=-1$ and $\mathcal{R}+\mathcal{T}=1$ (cf. Sec.~\ref{sec:non-hermitian-scattering}). This establishes a connection between the previous results for coherent impurities~\cite{Matveev1993,Yue1994,Nazarov2003} and dissipative ones~\cite{Froeml2019,Froeml2020}: from an open-system viewpoint, coherent impurities are a fine-tuned ($\gamma(x) \equiv 0$) and unstable special case within the more general class of complex impurities.
Finally, by identifying $\ell = -\log|k-k_\text{F}|$, we find, for a generic dissipative impurity, the following scaling for transmission and loss probability
\begin{equation}
	\mathcal{T} \sim \vert k-k_\text{F} \vert^{2\alpha}, \quad \eta \sim \vert k-k_\text{F} \vert^\alpha,
\end{equation} 
again confirming the result of Refs.~\onlinecite{Froeml2019,Froeml2020}.
In summary, for repulsive interactions, the only qualitative difference in the scaling of the probabilities is due to the presence of dissipation. The width of the barriers does not affect the scaling of the scattering probabilities. For that reason, interactions render particles with $k \rightarrow k_\text{F}$ dissipation-free, restoring the height of the Fermi edge.

\subsection{Stable fixed points for attractive interactions}

The line of fixed points FPL$_1$ is stable for attractive interactions ($\alpha<0$) and characterized by a continuum of values of  $\mathcal{T}^\star$. The first major implication of this is that, at the fixed point, the loss probability may acquire a finite value $\eta^\star >0$. This is in stark contrast to the case of a dissipative delta-shaped impurity studied in Refs.~\onlinecite{Froeml2019,Froeml2020}, where $\eta^\star=0$ for attractive interactions. This is a direct consequence of the existence of the quantity $\mathcal{D}$ conserved by the RG flow, see Eq.~\eqref{eq:Ddefinition}, whose value is fixed by the initial conditions, i.e., by the microscopic value of the scattering coefficients. In particular, at the line of fixed points FPL$_1$, the value of $\mathcal{D}$ can be expressed as
\begin{equation}
	\mathcal{D}=\frac{\eta^{\star 2}}{1-\eta^\star}.
\end{equation}
A straightforward computation using Eq.~\eqref{eq:Ddefinition} shows that in the case of delta-shaped impurities $\mathcal{D} = 0$, implying that $\eta^\star=0$, confirming the result of Refs.~\onlinecite{Froeml2019,Froeml2020}. For a generic dissipative impurity, instead, $\mathcal{D} \neq 0$, necessarily implying that $\eta^\star \neq 0$: accordingly, the finite width of a dissipative impurity qualitatively changes the fixed point of the RG flow.
We further investigate the stability properties of these fixed points by linearizing the RG equations around the fixed point, obtaining
\begin{equation}
	\partial_\ell 
	\begin{pmatrix}
	\delta\mathcal{R} \\ 
	\delta\mathcal{T} \\ 
	\delta \mathcal{X}
	\end{pmatrix}
	=
	 \alpha 
	 \begin{pmatrix}
	1-\mathcal{T}^\star & 0 & 0 \\ 
	-2\mathcal{T}^\star & 0 & 0 \\ 
	0 & 0 & 2\mathcal{T}^\star
	\end{pmatrix}
	\begin{pmatrix}
	\delta\mathcal{R} \\ 
	\delta\mathcal{T} \\ 
	\delta\mathcal{X}
	\end{pmatrix},
\end{equation}
which leads to the eigenvalues $0, 2\alpha\mathcal{T}^\star,\alpha(1-\mathcal{T}^\star)$: the marginal direction corresponds to the line of fixed points, while the other eigenvalues are negative, indicating stability of the line of fixed points. In contrast to the case of repulsive interactions, the eigenvalues do not depend only on $\alpha$, but also on the fixed-point value $\mathcal{T}^\star$.
Moreover, a stable direction becomes marginal when $\mathcal{T}^\star=1$: this corresponds to the delta-shaped impurities, which we discuss separately in App.~\ref{App:LocalizedImp}. Further insight can be gained by studying the scaling forms of $\delta\mathcal{T}, \delta\mathcal{R}$:
\begin{subequations}
	\begin{align}
		\delta\mathcal{R} &= \delta \mathcal{R}_0 e^{\alpha \eta^\star \ell}, \\
		\delta \mathcal{T} &= \delta\mathcal{T}_0 + \frac{2 \delta\mathcal{R}_0 \mathcal{T}^\star}{\eta^\star} - \frac{2 \delta\mathcal{R}_0 \mathcal{T}^\star}{\eta^\star} e^{\alpha \eta^\star \ell}. 
	\end{align}
\end{subequations}
As expected, a perturbation $\delta \mathcal{T}_0$ from a fixed point $\mathcal{T}^\star$ just displaces $\mathcal{T}$ to another fixed point $\mathcal{T}^{\star'} = \mathcal{T}^* + \delta\mathcal{T}_0 + 2 \delta\mathcal{R}_0 \mathcal{T}^\star/\eta^\star$. 
The scaling of $\delta\mathcal{R}$ is controlled by an exponent depending also on the fixed point value $\eta^\star = 1-\mathcal{T}^\star$.
This phenomenology is reminiscent of the Berezinskii–Kosterlitz–Thouless (BKT) transition~\cite{Kosterlitz:1973xp}, whose RG flow is characterized by a line of fixed points associated with a marginal operator, terminating at a point corresponding to the transition between a gapped and a gapless phase. The existence of a more substantial connection between the BKT physics and the one considered here is left open for future investigations. 

\begin{figure}
    \centering
    \includegraphics[width=0.35\linewidth]{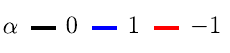}
    \includegraphics[width=\linewidth]{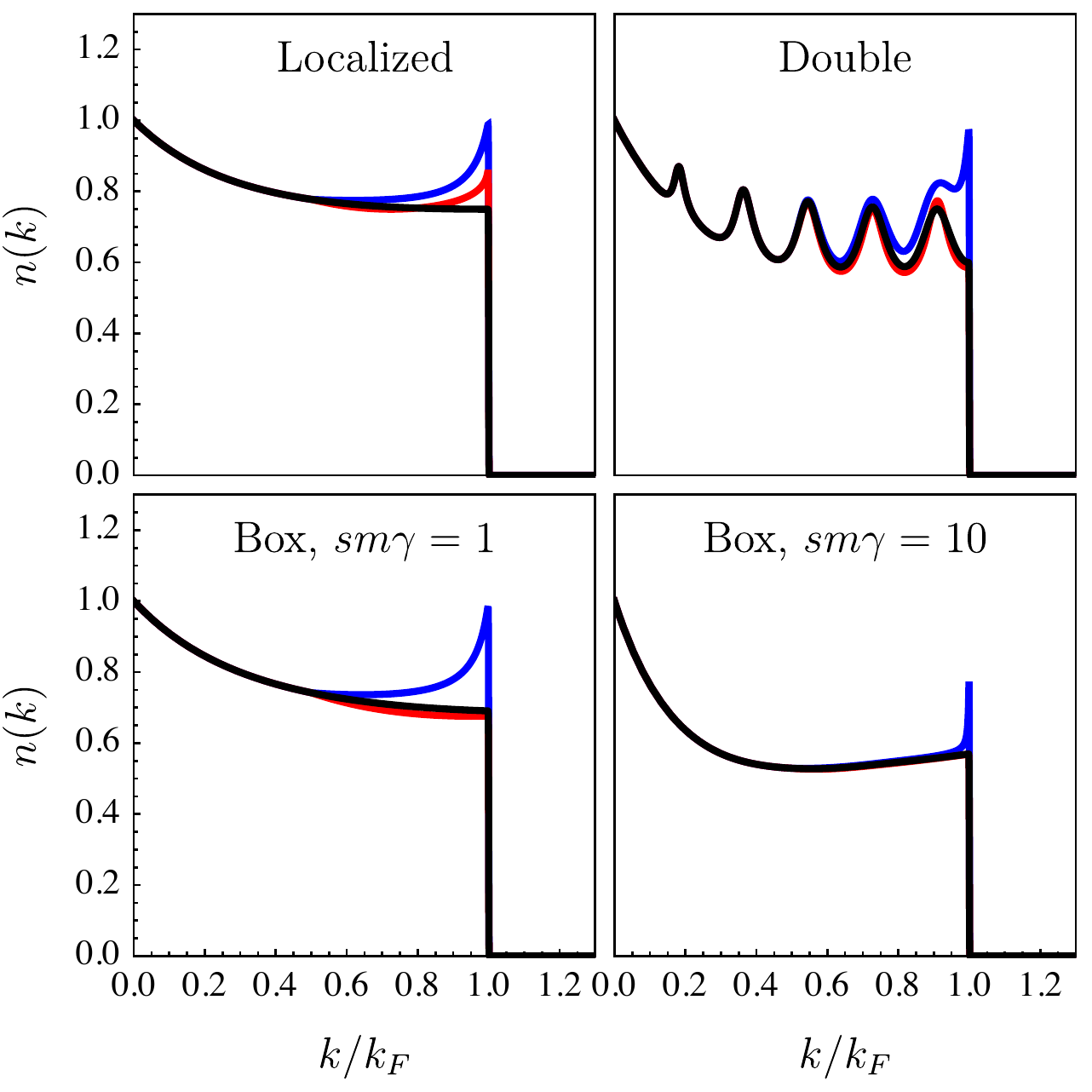}
    \caption{Momentum distribution for different interaction strengths and impurity shapes. For all panels $\Delta_\text{UV}=k_\text{F}/2,k_\text{F}=m \gamma$.}
    \label{fig:MomDistInter}
\end{figure}

Finally, the existence of fixed points with $\eta^\star \neq0$ for attractive interactions, but not for repulsive ones, can be physically rationalized as follows. In the first case, the reflection probability $\mathcal{R}$ is renormalized to zero, meaning that a particle impinging on the dissipative impurity is either transmitted or lost. If the impurity is delta-shaped, the renormalized wave function, i.e. the wave function with renormalized scattering parameters, develops a node at the point of the impurity, causing it to have vanishing overlap with the losses, and thus realizing perfect transmission (cf. Sec.~\ref{SubSec:resonantsup}). However for a dissipative impurity with a finite width, the wave function cannot avoid overlap with the losses by developing a node: accordingly, a finite amount of losses occurs. In the case of repulsive interactions, a particle impinging on the impurity is completely reflected by the effective single-particle potential generated by the Friedel oscillations: as a consequence, the wave function has vanishing overlap with the lossy impurity, regardless of its width. 
Attractive interactions therefore modify the Fermi edge in a non-universal way, instead of restoring it completely (cf. Fig.~\ref{fig:MomDistInter}, red curves).
The renormalization of the scattering parameters leads to a modification of the momentum distribution at the Fermi edge. Repulsive interactions ($\alpha>0$) always enhance the height of the Fermi edge to $1$, while attractive interactions modify it in a non-universal way ($\mathcal{\eta}^\star \neq 0$). Besides modifying the momentum distribution in absence of interactions, a finite impurity width also modifies the range of momenta where interactions take effect, acting as a UV cutoff.

\subsection{Unstable lines of fixed points}
Beyond the previously discussed lines of fixed points, there are two more, generically unstable lines of fixed points, FPL$_2$ and FPL$_4$. We discuss them here for completeness.

The flow becomes stationary for $\mathcal{R}=0$, $\mathcal{X}=-1$. Solving the linearized flow equations with initial values $\mathcal{R}_0,\mathcal{T}_0,\mathcal{X}_0$ yields
\begin{subequations}
	\begin{align}
		\mathcal{R} &= \mathcal{R}_0 e^{\alpha \left(1+\mathcal{T}^\star \right) \ell}, \\
		\mathcal{X} &= -1 + \left( 1+ \mathcal{X}_0 \right) e^{-2 \alpha \mathcal{T}^\star \ell},
	\end{align}
\end{subequations}
while $\mathcal{T}$ is marginal. For both repulsive and attractive interactions, one coupling is relevant, making the line of fixed points unstable. In order to observe the qualitatively modified scaling of $\mathcal{R}$ at FPL$_2$ compared to FPL$_1$, fine tuning to either $\mathcal{R}=0$ or $\mathcal{X}=1$ is necessary. The latter is for instance realized in the case of purely coherent impurities.

The remaining fixed points are found at $\mathcal{R}=\mathcal{T}=0$, i.e. $\eta=1$. In their vicinity, the leading order flow equations are
	\begin{subequations}
	\begin{align}
		\partial_\ell \mathcal{R} &=  \alpha \mathcal{R} , \\
		\partial_\ell \mathcal{T} &= -2 \alpha  \mathcal{R} \mathcal{T}.
	\end{align}
\end{subequations}
As both $\mathcal{R} \geq 0$ and $\mathcal{T} \geq 0$, one of the two parameters is enhanced by the RG flow, while the other one is decreased and this line of fixed points is again unstable. To observe its effect, a fine tuning to either $\mathcal{R}=0$ or $\mathcal{T}=0$ is needed, which is not found, even for resonant dissipative impurities (cf. Sec.~\ref{sec:ResonantDissipation}). Proximity to this line of fixed points further more enforces $\mathcal{D} \rightarrow \infty$, associated with an impurity of infinite width. This is clearly outside of the realm of our model.

\section{Conclusion and outlook}
\label{sec:conclusion}

In this article, we analyzed the effects of the shape of a dissipative impurity in a fermionic wire. First, we showed that multiple dissipative impurities generate resonant effects in the transmission and loss probabilities, in the latter case resulting in a clear signature in the particles' momentum distribution. Second, we found that the impurity's shape affects the scaling behavior of the scattering probabilities in the presence of interactions, giving rise to a novel line of fixed points in the RG equations describing the scaling. For repulsive interactions, the fluctuation-induced quantum Zeno effect is preserved. For attractive interactions, instead, the novel fixed points result in a vanishing reflection at the Fermi energy, accompanied by a finite loss, in contrast to the case of an ideal delta-shaped impurity.  

Our work opens several perspectives for future research. On the one hand, we showed that localized losses, a very controllable experimental tool, can be used to design spatial structures in ultracold gases, with properties similar to external potentials. Accordingly, localized losses may provide a valuable resource to quantum simulate novel states of matter, e.g.,  by designing non-Hermitian lattice structures~\cite{Ueda_review} or engineering the momentum distribution of ultracold particles.

On the other hand, a natural question concerns the analogies outlined with the BKT phase transition and, more generally, if non-Hermitian perturbations on equilibrium fixed points can generate nonequilibrium universal behavior with no counterpart at thermal equilibrium.

\begin{acknowledgments}
We acknowledge support by the funding from
the European Research Council (ERC) under the Horizon
2020 research and innovation program, Grant Agreement
No. 647434 (DOQS), by
the funding from the Deutsche Forschungsgemeinschaft
(German Research Foundation) under Germany’s
Excellence Strategy–Cluster of Excellence Matter and
Light for Quantum Computing (ML4Q) EXC 2004/1–
390534769, by the Deutsche Forschungsgemeinschaft
Collaborative Research Center (CRC) 1238 Project No.
277146847–project C04. M.~G. acknowledges funding  from  the  International  Max  Planck  Research School  for Quantum  Science  and  Technology  (IMPRS-QST).

\end{acknowledgments}

\appendix

\section{Keldysh field theory description} \label{App:KeldyshFieldTheory}

The dynamics of the density operator as given in Eq.~\eqref{QME} can be described using the Keldysh field theory for open quantum systems~\cite{1512.00637v2}. The corresponding Keldysh partition function reads:
\begin{subequations}
\begin{align}
	\mathcal{Z} &= \int \mathcal{D} \bar{\psi}_{c} \mathcal{D} \psi_{c} \mathcal{D} \bar{\psi}_{q}\mathcal{D} \psi_{q} e^{i S[\bar{\psi}_{c},\psi_{c},\bar{\psi}_{q},\psi_{q}]}, \\
	S &= \int_{x,\omega} \left( \begin{array}{cc}
	\bar{\psi}_c & \bar{\psi}_q
	\end{array}  \right) \left( \begin{array}{cc}
	0 & P_\text{A} \\ 
	P_\text{R} & P_\text{K}
	\end{array}  \right) \left( \begin{array}{c}
	\psi_c \\ 
	\psi_q
	\end{array}  \right) + S_\text{int}.
\end{align}
\end{subequations}
All quantities are evaluated at frequency $\omega$ and position $x$. The bare retarded Green's function is given as the inverse of the operator
\begin{equation}
	P_\text{R} = \omega + i0^+ + \frac{\partial_x^2}{2m} - V(x) + i \gamma(x) ,
\end{equation}
$G_\text{R}=P_\text{R}^{-1}$. $0^+>0$ is a regularization ensuring causality. For a non-interacting system, a single particle picture is valid and described by a Schrödinger equation for particles moving forward in time,
\begin{equation}
	\omega \phi(x)= \left[ \frac{-\partial_x^2}{2m} - \left( V(x) - i \gamma(x) \right) \right] \phi(x),
\end{equation}
where $\phi(x)$ is a single particle wave function. This allows one to identify the complex potential $U(x)=V(x)-i \gamma(x)$ and the effective non-Hermitian single particle Hamiltonian $\mathcal{H}(x)=\frac{-\partial_x^2}{2m}+U(x)$ and treat the non-interacting dynamics of the system as a scattering problem with a non-Hermitian Hamiltonian as it is done in Sec.~\ref{sec:non-hermitian-scattering}. 

\section{Causality constraint}
\label{app:Causality}

The physically motivated constraint \eqref{CausalityCondition} implies in particular
\begin{equation}
	\max_{\vec{\psi}^\text{in} \in \mathbb{C}^2}   \frac{\vec{\psi}_k^{\text{in} \dagger} S_k^\dagger S_k \vec{\psi}_k^{\text{in}}}{\vec{\psi}_k^{\text{in} \dagger} \vec{\psi}_k^\text{in}} \leq 1.
\end{equation}
By a parametrization of the vector $\vec{\psi}^\text{in}=\left(1,a_k \right)^T$ with $a_k \in \mathbb{C}$, we find the equivalent condition
\begin{equation}
	\min_{a \in \mathbb{C}} \left( \eta^+_k + \vert a_k \vert \eta_k^- -2 \text{Re} \left[ a_k \left( r_k^{+*} t_k + r_k^- t_k^* \right) \right] \right) \geq 0.
\end{equation}
The minimization can be easily done and yields
\begin{equation}
	\vert r_k^{+*} t_k + r_k^- t_k^* \vert \leq \eta_k^+ \eta_k^-.
\end{equation}
In case of $\mathcal{R}_k^+ = \mathcal{R}_k^-$, we find the relation
\begin{equation}
	\eta^2_k -2 \mathcal{R}_k \mathcal{T}_k (1+\mathcal{X}_k) \geq 0
\end{equation}
from the main text.

\section{Scattering Green's function}
\label{app:ScatGreen}
Here, we show the structure of the scattering Green's function for a dissipative impurity of finite extent, using the Dyson equation. In the absence of the impurity, the retarded Green's function is translation-invariant $G_\text{R}(\omega,x,y)=G_\text{R}(\omega,x-y)$ and reads in a mixed position and momentum representation 
\begin{equation}
	G_\text{R}^0(\omega,x,k) = \int_y e^{iky} G_\text{R}(\omega,x,y)  =\frac{e^{ikx}}{\omega+i0^+-k^2/(2m)}. 
\end{equation}
The causal structure is preserved when the impurity is included so that we parametrize the full Green's function as
\begin{equation}
	G_\text{R}^0(\omega,x,k) = \frac{f(\omega,x,k)}{\omega+i0^+-k^2/(2m)},
\end{equation}
with a function $f$ without poles. The impurity is then included exactly by solving a Dyson equation for this function:
\begin{equation} \label{ffunc}
	f(\omega,x,k) = e^{ikx} + \frac{1}{iv_\omega} \int_y U(y) f(\omega,y,k) e^{ik_\omega \vert x-y \vert}.
\end{equation}
Here, we introduced the velocity $v_\omega = \sqrt{2\omega /m}$ and the momentum $k_\omega = \sqrt{2m \omega}$. For a localized impurity, this can be solved explicitly while we take Eq.~\eqref{ffunc} as a definition for finite width impurities.\\
Considering again the definition of the retarded Green's function, we obtain
\begin{align} \label{Eq:WaveAndGreen}
	\left( \omega - H(x) \right) G_\text{R}(\omega,x,y) &= \delta(x-y) \notag \\
	\Rightarrow \left( \omega - H(x) \right) f(\omega,x,k) &= \left( \omega-\frac{k^2}{2m} \right) e^{ikx} \notag \\
	\Rightarrow f(k^2/(2m),x, \pm k) &= \phi_k^\pm(x).
\end{align}
This allows us to relate the Green's function to the scattering wave functions from the main text (cf. Sec.~\ref{sec:non-hermitian-scattering}). \\
The pole structure of the scattering Green's function may also be used to simplify it in the limit of long times. Consider
\begin{align}
	&G_\text{R}(t,x,k) = \int_\omega  \frac{f(\omega,x,k) e^{-i \omega t}}{\omega+i0^+-k^2/(2m)} \notag \\
	&= -i \int_{\omega,t'} \Theta(t+t') e^{i\left(\omega-k^2/(2m) \right)t'} f(\omega,x,k) e^{-itk^2/(2m)}  .
\end{align}
In the limit $t \rightarrow \infty$, we have $\Theta(t+t')=1$ such that the Green's function simplifies to
\begin{equation} \label{Eq:LongTime}
	G_\text{R}(t \rightarrow \infty,x,k) = -if(k^2/(2m),x,k) e^{-itk^2/(2m)}.
\end{equation}
The scattering wave function appears explicitly here. This long-time limit is used in App.~\ref{app:StationaryState} to compute stationary state properties of the system.

\section{Stationary state density}
\label{app:StationaryState}

Here, we sketch the derivation of the stationary state distribution $C(x,y)$ as defined in Eq.~\eqref{Eq:Corrfunc} in the absence of interactions. We use the Heisenberg picture and define time-dependent operators $\hat{\psi}^\dagger(t,x)$ and $\hat{\psi}(t,x)$ that obey the condition
\begin{align}
	\text{tr} \left[ \hat{\rho} \partial_t \left( \hat{\psi}^\dagger(t,x) \hat{\psi}(t,y) \right) \right]  = \text{tr} \left[ \hat{\psi}^\dagger(x) \hat{\psi}(y)   \partial_t \hat{\rho}(t) \right].
\end{align}
The temporal evolution of the density operator $\hat{\rho}(t)$ is given by the quantum master equation~\eqref{QME}. Using the cyclic property of the trace yields again an equation for expectation values that is solved if the time-dependent fermionic operators fulfill
\begin{subequations}
\begin{align}
	\partial_t \hat{\psi}(t,x) &= -i\mathcal{H}(x) \hat{\psi}(t,x), \\
	\partial_t \hat{\psi}^\dagger(t,x) &= i\mathcal{H}^*(x) \hat{\psi}^\dagger(t,x).
\end{align}
\end{subequations}
This is solved for $t>0$, after the quench of the impurity, by
\begin{subequations}
\begin{align}
	\hat{\psi}(t,x)&= \int_{x'} G_\text{R}(t,x,x') \hat{\psi}(0,x'), \\
	\hat{\psi}^\dagger(t,x)&= \int_{x'} G_\text{R}^*(t,x,x') \hat{\psi}^\dagger(0,x'),
\end{align}
\end{subequations}
where $G_\text{R}=P_\text{R}^{-1}$ is the retarded Green's function of the non-interacting system including the complex potential of the impurity. Hence, the time-dependent correlation function is given by
\begin{equation}
	C_t(x,y) = \int_{x',y'} G_\text{R}^*(t,x,x') G_\text{R}(t,y,y') C_0(x',y').
\end{equation}
Before the quench of the impurity, we assume the system to be in its ground state
\begin{equation}
	C_0(x,y)=\left\langle \hat{\psi}^\dagger(x) \hat{\psi}(y) \right\rangle_{\hat{\rho}_0} = \int_k e^{-ik(x-y)} n_\text{F}(k),
\end{equation}
where $n_\text{F}(k)=\Theta(k_\text{F}^2-k^2)$ is the Fermi distribution with the Fermi momentum $k_\text{F}$. This yields
\begin{equation}
	C_t(x,y)=\int_{-k_\text{F}}^{k_\text{F}} \frac{dk}{2\pi} G_\text{R}^*(t,x,k) G_\text{R}(t,y,k) .
\end{equation}
We are now interested in the stationary state properties of the system long after the quench of the impurity. Using Eq.~\eqref{Eq:LongTime} and~\eqref{Eq:WaveAndGreen}, we find Eq.~\eqref{eq:correlation} from the main text.

\section{Momentum distribution}
\label{App:MomentumDistribution}

Here we sketch the derivation of the momentum distribution in the stationary state as defined in Eq.~\eqref{Eq:Mom-Dist-Def}. For that purpose, we write the wave function as
\begin{equation}
	\phi_q(x) = e^{iqx} + \begin{cases}  A_q e^{-i \vert q \vert x} & x < 0 \\ B_q e^{i \vert q \vert x} & x > 0 \end{cases}.
\end{equation}
Where we neglected the impurity-region as it contributes only $ \sim s/L$ to the momentum distribution and introduced $A_{\vert q \vert}=r^+_{\vert q \vert}, A_{-\vert q \vert}=t_{\vert q \vert}-1$ and $B_{-\vert q \vert}=r^-_{\vert q \vert}, B_{\vert q \vert}=t_{\vert q \vert}-1$ for brevity.  The momentum distribution becomes
\begin{subequations}
\begin{align}
	n(k) &= \frac{1}{L} \int_{x,y} e^{ik(x-y)} \int_q n_\text{F}(q) \phi_q^*(x) \phi_q(y) \\
	&= \frac{1}{L} \int_q n_\text{F}(q) \left\vert  \phi_q(k) \right\vert^2. 
\end{align}
\end{subequations}
For the Fourier transformation of the wave function, we use a regularization for long distances, $e^{-\vert x \vert/L}$, and find
\begin{equation}
	\phi_q(k) = 2 \pi \delta(k-q) + \frac{iA_q}{k + \vert q \vert + i/L} + \frac{iB_q}{-k+\vert q \vert + i/L} . 
\end{equation}
We use $\int_x = 2\pi \delta(k=0)=L$ and obtain in the limit $L \rightarrow \infty$
\begin{equation}
	n(k) = n_\text{F}(k) \begin{cases} 1+\text{Re } A_k + \frac{1}{2} \vert A_k \vert^2 + \frac{1}{2} \vert A_{-k} \vert^2 & k<0 \\ 1+\text{Re } B_k + \frac{1}{2} \vert B_k \vert^2 + \frac{1}{2} \vert B_{-k} \vert^2 & k<0  \end{cases}.
\end{equation}
In terms of the scattering parameters, this gives Eq.~\eqref{Eq:Mom-Dist-Res} from the main text.

\section{Symmetry of transmission parameters under inversion}
\label{app:TransferMatrix}

Here we discuss the statements about symmetries of scattering parameters as made in Sec. \ref{sec:non-hermitian-scattering} in more detail. In Hermitian scattering problems, the unitarity of the scattering matrix $S_k^\dagger S_k=\mathds{1}$ holds due to the conservation of probability. If the Schrödinger equation is in addition time-reversal invariant, $S_k$ is symmetric $S_k^T = S_k$ such that the transmission parameters have to fulfill $t_k^+=t_k^-$ for arbitrary impurities. In particular, this holds then for non-inversion-symmetric profiles. \\
Both used symmetries, quantum-mechanical time-reversal and conservation of probability, are broken in the case of dissipative impurities. Still, $t_k^+=t_k^-$ holds, even for inversion-asymmetric impurities, as we now demonstrate. Consider a sufficiently smooth and compact complex potential $U(x)$ such that it can be approximated to arbitrary accuracy to be piecewise constant 
\begin{equation}
	U(x) = \begin{cases} 0 & x<-s/2 \\ U_1 & -s/2 < x< x_1 \\ \vdots & \vdots \\
	U_N & x_{N-1}<x<s/2 \\ 0 & x>s/2 \end{cases}.
\end{equation}
We introduced positions $x_i$ separating segments of constant (complex) potential $U_i$. Using the transfer matrix method~\cite{TransferMatrix}, we may compute the scattering parameters of the compound impurity from the scattering parameters of every single constant potential segment as $M_k = \prod_{i=1}^N M_{k,i}$ where $M_k$ is the transfer matrix of the compound impurity and $M_{k,i}$ is the transfer matrix of the segment from $x_{i-1}$ to $x_i$. Every single segment is inversion-symmetric about its center such that we know $t_{k,i}^+=t_{k,i}^-$ for the transmission parameters of such a segment. Equivalently, we find $t_{k,i}^+/t_{k,i}^-=\det M_{k,i}=1$ such that we get for the total scattering parameters
\begin{equation}
	\frac{t_k^+}{t_k^-} = \det M_k = \det \prod_{i=1}^N M_{k,i} = \prod_{i=1}^N \det M_{k,i}=\prod_{i=1}^N 1 = 1 .
\end{equation}  
Hence, it is not possible to construct an impurity with asymmetric transmission properties $t_k^+ \neq t_k^-$ out of inversion-symmetric segments such that we may always assume $t_k^+=t_k^-=t_k$ in the main text.  Such an argument does not exist for the reflection parameters and there are indeed inversion-asymmetric impurities, where both phase and amplitude of $r_k^+$ and $r_k^-$ differ (cf. Sec.~\ref{sec:ResonantDissipation}).

\section{Dissipative impurity close to a hard wall}
\label{App:Wall}

Here we consider another physically relevant situation, beyond the scope of the main text, where a dissipative resonance arises: a lossy impurity close to a hard wall, i.e., to a point where the wave function vanishes. To this end, we take the problem of a semi-infinite wire, extending from $x \to -\infty$ to $x=0$. A loss impurity with strength $\gamma$ is then located at $x=-D$. In this setting, the scattering properties are completely described by the reflection amplitude $r_k$ which fixes $\eta_k=1-\vert r_k \vert^2$.

\begin{figure}
\centering
\includegraphics[width=\linewidth]{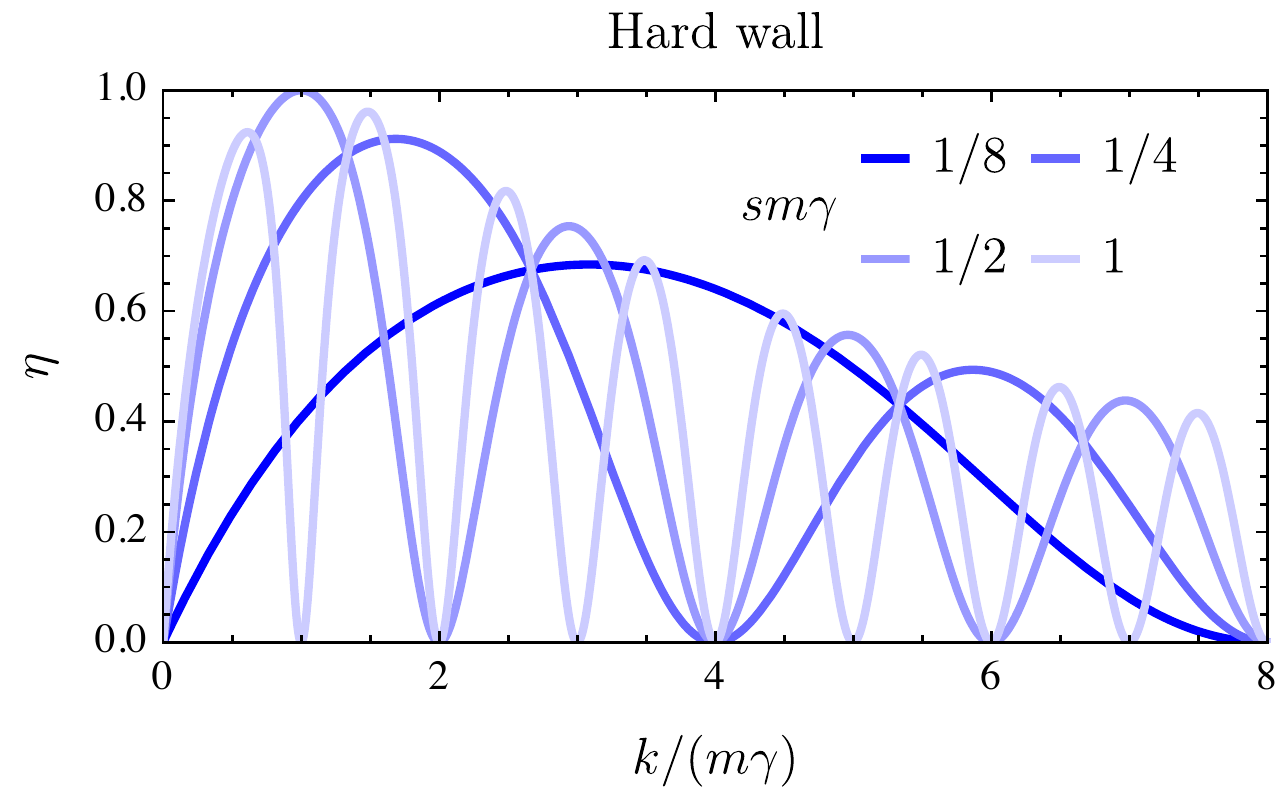}
\caption{Loss probability for a localized dissipation near the end of a wire. Depending on the distance $s$, the dissipation is strongly enhanced or perfectly removed. For $k=m \gamma$, the loss probability can reach $1$ perfectly.}
\label{Fig:WireEnd}
\end{figure}

The presence of the infinite barrier imposes a strict von Neumann boundary condition $\phi(0)=0$, leaving the ratio between loss and reflection probability still to be determined by the additional loss. However, this boundary condition implies that the wave function vanishes at $x=-\pi n/ \vert k \vert$ for $n \in \mathbb{N}$. If the localized impurity is at such a point, it is not seen by an incoming particle of momentum $k$ which is therefore always reflected. Away from the nodes, the loss probability of this setup is enhanced as particles that are reflected from the boundary cross the region with dissipation more than once. 
This gives rise to a momentum-comb structure in the loss probability (cf. Fig.~\ref{Fig:WireEnd}). 
In this setup, the RG method from Sec.~\ref{Sec:RG} is applicable, even though the loss probability is highly asymmetric: due to the exclusion of the region $x>0$ by the potential, the effect of interactions has to be included only for $x<0$. Hence, the ansatz for the scattering wave functions is not modified by interactions and we may apply the same strategy as discussed previously even though the impurity is not inversion-symmetric. We find a single RG flow equation for the loss probability
\begin{equation}
	\partial_\ell \eta_k = -\alpha \eta_k (1-\eta_k).
\end{equation}
This means in particular that attractive interactions enhance the dissipation to perfect particle loss with perfect reflection being an unstable fixed point. The possibility to reach this unstable fixed point by tuning $D$ gives rise to the competition between resonant reflection due to the impurity shape and perfect loss due to interactions.

\section{Localized impurities as limiting case}
\label{App:LocalizedImp}
\begin{figure}
\includegraphics[width=\linewidth]{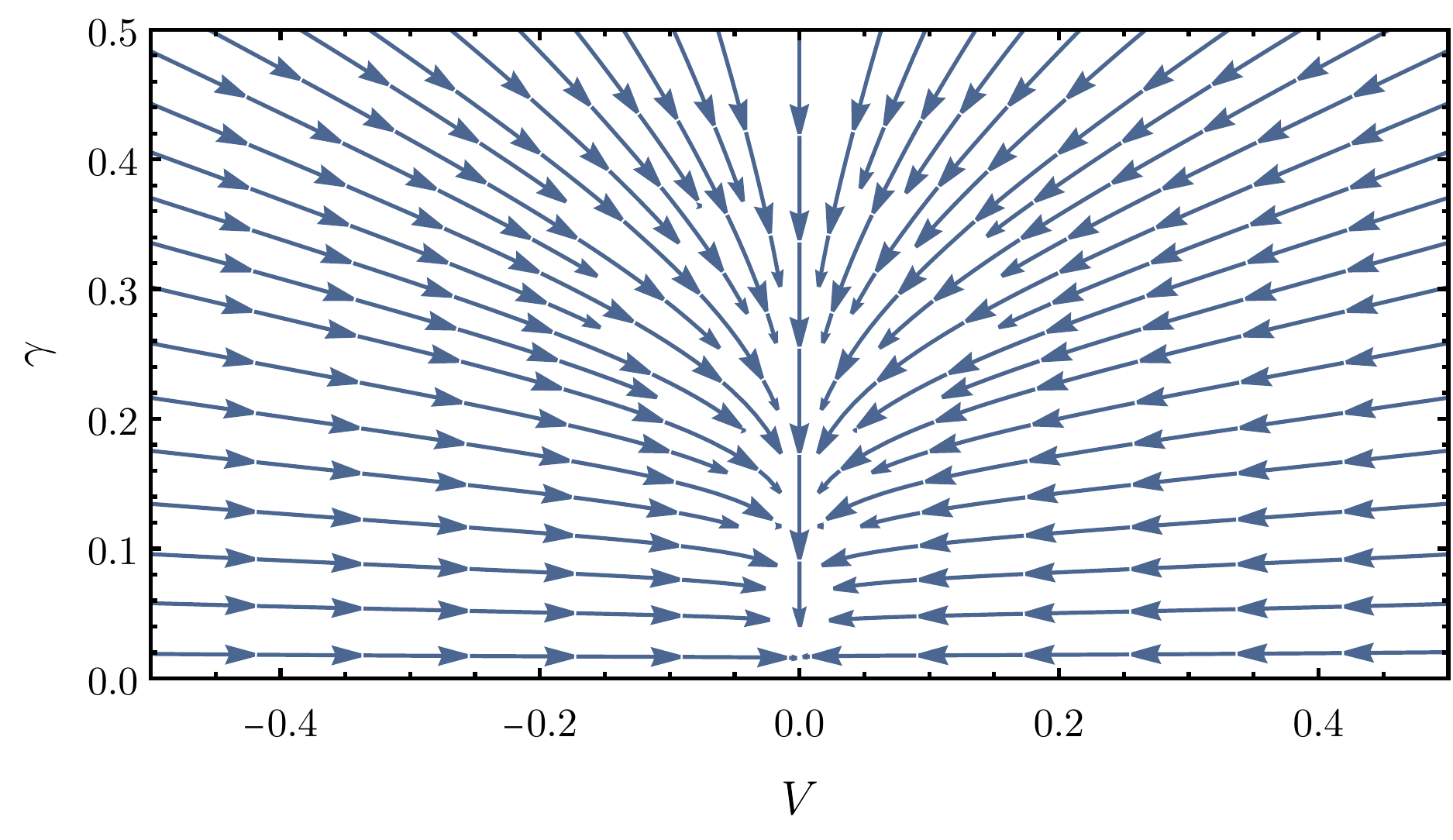}
\caption{Renormalization of the effective strength of coherent and dissipative localized impurities for attractive interactions. The stable fixed point lies at $\gamma= V=0$ corresponding to perfect transmission. $V$ vanishes algebraically while $\gamma$ vanishes only logarithmically slow so that the dissipative part dominates close to the Fermi edge.}
\label{2DFlow}
\end{figure}
The appearance of the line of fixed points along $\mathcal{X}$ has a simple explanation in terms of the extremal case of localized impurities,
\begin{equation}
	U(x)=\left(V-i \gamma \right) \delta(x),
\end{equation} 
where $V \in \mathbb{R}$ and $\gamma>0$ are the coherent and dissipative component of the impurity, respectively.
These impurities fulfill the condition $\mathcal{D}=0$ and the corresponding surface is fully parametrized by the real and imaginary components, $V$ and $\gamma$. Because of this, we can describe all RG flow trajectories connected to the point $\mathcal{T}^\star=\mathcal{X}^\star=1,\mathcal{R}^\star =0$, where the scaling exponent vanishes, by the two parameters $V$ and $\gamma$ of such an impurity. The flow equations then become (with $v_k=1$)
\begin{subequations}
\begin{align}
	\partial_\ell V &= \alpha V \left( 1- \frac{\gamma}{(1+\gamma)^2 + V^2} \right), \\
	\partial_\ell \gamma &= \alpha \gamma \left( 1-  \frac{1+\gamma}{(1+\gamma)^2 + V^2} \right).
\end{align}
\end{subequations}
This RG flow becomes stationary for $\gamma=V=0$ and $\gamma^2 + V^2 \rightarrow \infty$. The latter is attractive for repulsive interactions, and the two equations decouple to $\partial_\ell \gamma = \alpha \gamma$ and $\partial_\ell V = \alpha V$ such that both parameters grow $\sim \Delta_\text{IR}^\alpha$. Because of this, the ratio $\gamma/V$ approaches a non-universal constant that fixes the non-universal $\mathcal{X}^\star$. In this way, we can interpret the marginal coupling $\mathcal{X}$ of the general flow equations as a ratio of two equally relevant couplings. The perfect reflectivity in this case is understood as the effective strength of the impurity is increased to infinity due to the interactions. It is an incarnation of the quantum Zeno effect that an infinitely strong dissipation acts exactly like a strong barrier, and transmission is suppressed \cite{Froeml2019,Froeml2020}. \\
In the opposite case of attractive interactions, the fixed point of vanishing effective impurity strength becomes attractive and the flow equations decouple. While the coherent part simplifies again to $\partial_\ell V = \alpha V$ with the same algebraic scaling as in the strong impurity limit, the leading order in $\gamma$ is quadratic, $\partial_\ell \gamma = \alpha \gamma^2/2$. This means that the dissipation strength scales logarithmically,
\begin{equation}
	\gamma \sim \frac{1}{\alpha \log \Delta_\text{IR} / \Delta_\text{UV}}.
\end{equation} 
Because of this, at $\Delta_\text{IR} \rightarrow 0$, the dissipation dominates over the coherent part so that the result of a purely dissipative impurity is reproduced. The logarithmic scaling of $\gamma$ is also translated to a logarithmic scaling of the corresponding reflection and loss probability
\begin{equation}
	\mathcal{R} \sim \frac{1}{( \alpha \log \Delta_\text{IR}/\Delta_\text{UV})^2}, \quad \mathcal{\eta} \sim \frac{1}{ \alpha \log \Delta_\text{IR}/\Delta_\text{UV}}.
\end{equation}
\\
This logarithmic behavior presented in Refs. \onlinecite{Froeml2019,Froeml2020} therefore replaces the algebraic scaling discussed in Sec.~\ref{Sec:ScalingGeneric} in the limiting case of $\mathcal{\eta}^\star=0$, where the scaling exponent vanishes.

\bibliography{Bibliography.bib}

\end{document}